\documentclass[aps,pra,twocolumn,longbibliography,floatfix,superscriptaddress]{revtex4-1}
\usepackage{amsfonts}
\usepackage{amssymb}
\usepackage{amsmath}
\usepackage{amsthm}
\usepackage{bm}% bold math
\usepackage{braket}

\usepackage{graphicx}
\usepackage{dcolumn}% Align table columns on decimal point

\usepackage[T1]{fontenc}
\usepackage[latin9]{inputenc}
\usepackage[normalem]{ulem}
\usepackage{xcolor}
\usepackage[breaklinks=true]{hyperref}
\hypersetup{colorlinks=true,linkcolor=blue,citecolor=blue,filecolor=blue,urlcolor=blue,pdfstartview=FitH}

\definecolor{OliveGreen}{RGB}{0,100,0}

\def\ex{\mathbf{e}_x}  
\def\ey{\mathbf{e}_y}  
\def\ez{\mathbf{e}_z}  

\begin{document}
\title{Dynamical Instability of 3d Stationary and Traveling Planar Dark Solitons}

\author{T.~Mithun}
\affiliation{Department of Mathematics and Statistics, University of Massachusetts, Amherst MA 01003-4515, USA}

\author{A.~R.~Fritsch}
\affiliation{Joint Quantum Institute, National Institute of Standards and Technology, and University of Maryland, Gaithersburg, Maryland, 20899, USA} 

\author{I.~B.~Spielman}
\email{ian.spielman@nist.gov}
\affiliation{Joint Quantum Institute, National Institute of Standards and Technology, and University of Maryland, Gaithersburg, Maryland, 20899, USA} 

\author{P.~G.~Kevrekidis}
\email{kevrekid@umass.edu}
\affiliation{Department of Mathematics and Statistics, University of Massachusetts, Amherst MA 01003-4515, USA}

\begin{abstract}
Here we revisit the topic of stationary and propagating solitonic excitations in self-repulsive  three-dimensional Bose-Einstein condensates by quantitatively comparing theoretical analysis and associated numerical computations with our experimental results.  
Using fully 3d numerical simulations, we explore the existence, stability, and evolution dynamics of planar dark solitons, as well as their instability-induced decay products including solitonic vortices and vortex rings. 
In the trapped case and with no adjustable parameters, our numerical findings are in correspondence with experimentally observed coherent structures.
Without a longitudinal trap, we identify numerically exact traveling solutions and quantify how their transverse destabilization threshold changes as a function of the solitary wave speed. 
\end{abstract}

\maketitle

% -------------------------------
\section{Introduction}
Dark solitonic structures emerge ubiquitously in systems that combine dispersion with a self-defocusing nonlinearity.
Numerous physical examples encompassing these properties prominently include atomic Bose-Einstein condensates (BECs)~\cite{kevrekidis2015defocusing,Frantzeskakis_2010}, and nonlinear Kerr (and generalized non-Kerr) media~\cite{KIVSHAR199881}. 
More broadly, solitonic structures also arise in mechanical lattices of coupled pendula~\cite{PhysRevLett.68.1730}, electrical transmission lines~\cite{PhysRevE.49.828}, microwave magnetic thin films~\cite{PhysRevLett.70.1707,PhysRevLett.84.4697}, acoustic waveguides of Helmholtz resonators~\cite{PhysRevE.91.023204}, surface water waves~\cite{PhysRevLett.110.124101}, and nematic liquid crystals~\cite{Piccardi:11}.
Outside of these closed systems, solitonic waves are even present in media balancing gain and loss, such as polariton fluids in semiconductor microcavities~\cite{carusotto_2013}. The presence of solitary waves in such disparate contexts attests to the breadth of their relevance and the generality of the central mechanisms leading to their creation. In particular, atomic BECs hold a central role amongst these systems owing to their highly tunable interaction Hamiltonians, the control of dimensionality and potential landscapes, as well as the diverse available mechanisms for creating excitations~\cite{pethick,stringari,kevrekidis2015defocusing}.

There are different ways BECs are used to create solitonic states. For example, dragging a laser beam through a BEC~\cite{PhysRevLett.99.160405}, phase-imprinting~\cite{Becker2008}, and matter-wave interference~\cite{PhysRevLett.101.130401,PhysRevA.81.063604}. More recent experiments showed control of both the position and velocity of soliton~\cite{Fritsch2020creating}, with contemporary theoretical studies exploring 3-dimensional (3d) solitary waves in detail~\cite{mateo2015stability}. The intriguing dynamics of the resulting structures, including their breakup into vortical patterns~\cite{PhysRevLett.86.2926,Shomroni2009} (even in superfluid Fermi gases~\cite{PhysRevLett.113.065301}) is a subject of wide investigation and has a wide variety of potential applications ranging from atomic matter-wave interferometers~\cite{Negretti_2004} to producing two-level qubit systems~\cite{PhysRevA.95.053618}. Much of this experimental and theoretical effort is summarized in Refs.~\cite{Frantzeskakis_2010,kevrekidis2015defocusing}.

In the present work, we revisit dark solitary waves 3d repulsively interacting BECs highly elongated along the longitudinal axes $\ez$. These waves are associated with a reduction in the local atomic density accompanied with a modification of the condensate phase. In elongated systems, vortex rings have a similar density distribution but with a ring-shaped phase singularity~\cite{komineas,kevrekidis2015defocusing}. The simplest example is a planar (or kink) soliton where the density is reduced to be two dimensional on a plane normal to the soliton's direction of propagation. By contrast, a solitonic vortex is a highly anisotropic vortex with a linear phase-singularity~\cite{kevrekidis2015defocusing}.  
%but with a density distribution reminiscent of a kink soliton~\cite{kevrekidis2015defocusing}. 
The potential conversion of kink solitons into these excitations is of wide interest~\cite{PhysRevA.68.043617,PhysRevA.65.043612,Becker_2013,tylutki2015solitonic,PhysRevA.94.023627,PhysRevLett.113.065302}.

We begin by comparing experimentally observed solitary waves with 3d numerical simulations computed with no adjustable parameters. Motivated
by this comparison, we theoretically explore the related 3d case of a longitudinally traveling solitonic wave in a system that is transversely confined but longitudinally infinite. Then, we compute the relevant Bogolyubov-de Gennes (BdG) modes to identify the onset of transverse instability and to quantify the growth rate of the unstable mode. This process confirms the prediction of Ref.~\cite{PhysRevLett.89.110401} that the interval of stability of kink solitons increases as their speed approaches the local speed of sound c. Finally in the unstable regime, we numerically follow the exponential growth of the unstable mode to its ultimate conversion to vortex structures.

\begin{figure*}[t]
\includegraphics{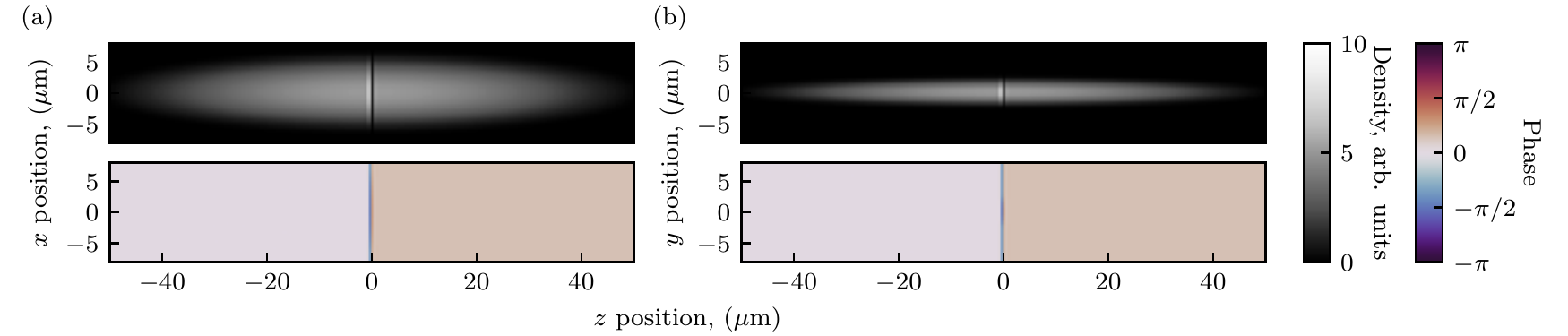}
	\caption{Integrated GPE wavefunction density (top) and phase cross-section (bottom) evaluated 
	%$1\ \mu\mathrm{s}$ 
	immediately after a $t_p = 140\ \mu\mathrm{s}$ phase imprint.
	(a) and (b) plot  $|\psi(x,z)|^2$ and $|\psi(y,z)|^2$ respectively.}
    \label{fig:PHE2D0}
\end{figure*}

\section{Experimental Motivation: Phase Engineering and In-Trap Motion}

In the present work, we model the experimental procedure in reference~\cite{Fritsch2020creating} which describes a new wavefunction engineering approach for launching solitons in atomic BECs and show that dark planar solitons are first created and then relax into a solitonic vortex.

The experiments started with $N=2.4\times10^5$ $^{87}\text{Rb}$ BECs confined in an optical dipole trap with frequencies $(\omega_x,\omega_y,\omega_z)=2\pi \times (94.73,153.23,9.1)\ \text{Hz}$ and chemical potential $\mu=h\times1.1$~kHz. 
We created solitonic excitations by combining
conventional phase imprinting with density engineering, i.e., wavefunction engineering, yielding solitonic excitations with a wide range of initial velocities~\cite{Fritsch2020creating}.

%In experiment, a phase imprinting potential $V_t=h\times5.5(3)$~kHz is applied to half of the extension of our elongated condensate for varied time $t_p$ to imprint a phase $\phi=V_t t_p/\hbar$.  
We consider the fully $3d$ Gross-Pitaveskii model (GPE), complementing these experiments~\cite{pethick,stringari}
\begin{equation}
  i\hbar\frac{\partial\psi}{\partial t}= \left[-\frac{\hbar^2}{2m}\nabla^2+V(\mathbf{r})+g|\psi|^2 \right] \psi.
  \label{eq:3DGP}
\end{equation}
Here, $\psi$ is normalized to the total atom number $N_{\mathrm{3D}}$,
interaction constant and potential are defined as $g=4 \pi \hbar^2 a_s/m$ and $V(\mathbf{r})=m(\omega_x^2 x^2+\omega_y^2 y^2+\omega_z^2 z^2)/2+V_\mathrm{step}(t) \mathcal{H}(z)$, respectively, where $\mathcal{H}(z)$ represents the Heaviside function.
We fix the magnitude of the step potential $V_\mathrm{step}(t)$ to be $0$ except during the phase-imprinting pulse when it is $h\times5.5\ \mathrm{kHz}$.

We simulate the experiments using the parameters described above and create solitons via the phase imprinting method: applying a potential generated with an infinitely sharp edge (smaller than the BEC healing length) to half of the BEC.
We note that the density engineering method is implemented in the experiment to circumvent technical limitations, given that the BEC healing length is typically much smaller than the optical resolution of the imaging systems used for phase imprinting, which is not the case in numerical simulations.
We found that the idealization of a step potential centered at $z=0$ and amplitude $V_{\mathrm{step}}$, is sufficient to reproduce the experimental results.
This indirectly demonstrates the efficacy of the experimental approach.

\begin{figure*}
     \includegraphics[width=\textwidth,angle =0]{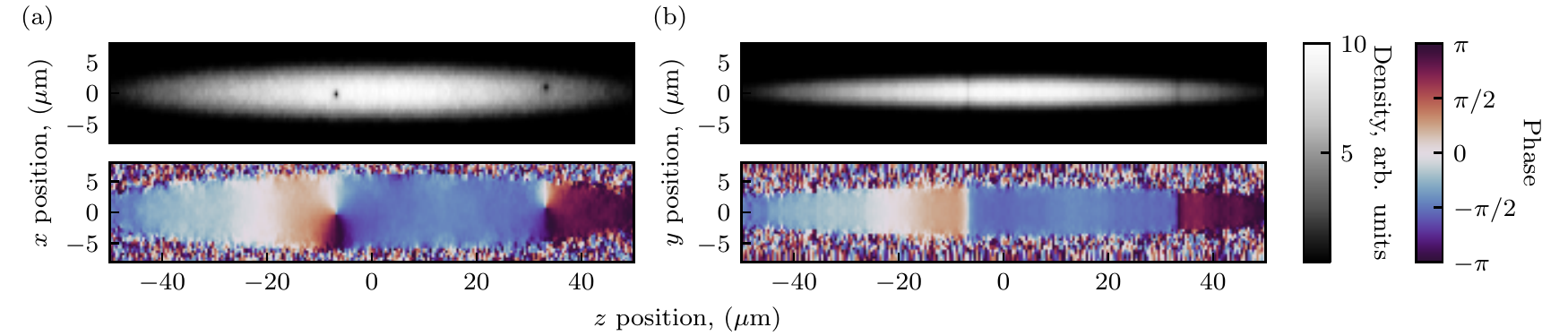}
       \caption {
Integrated GPE wavefunction density (top) along with phase (bottom) cross-sections evaluated $400\ \mathrm{ms}$ after a $t_p = 140\ \mu\mathrm{s}$ phase imprint.
(a) plots the $|\psi|^2$ projection and associated cross-sections in the $\ex-\ez$ plane and (b) plots $\ey-\ez$ quantities.
 }
    \label{fig:PHE2D}
\end{figure*}

\begin{figure}
     \includegraphics[width=\columnwidth,angle =0]{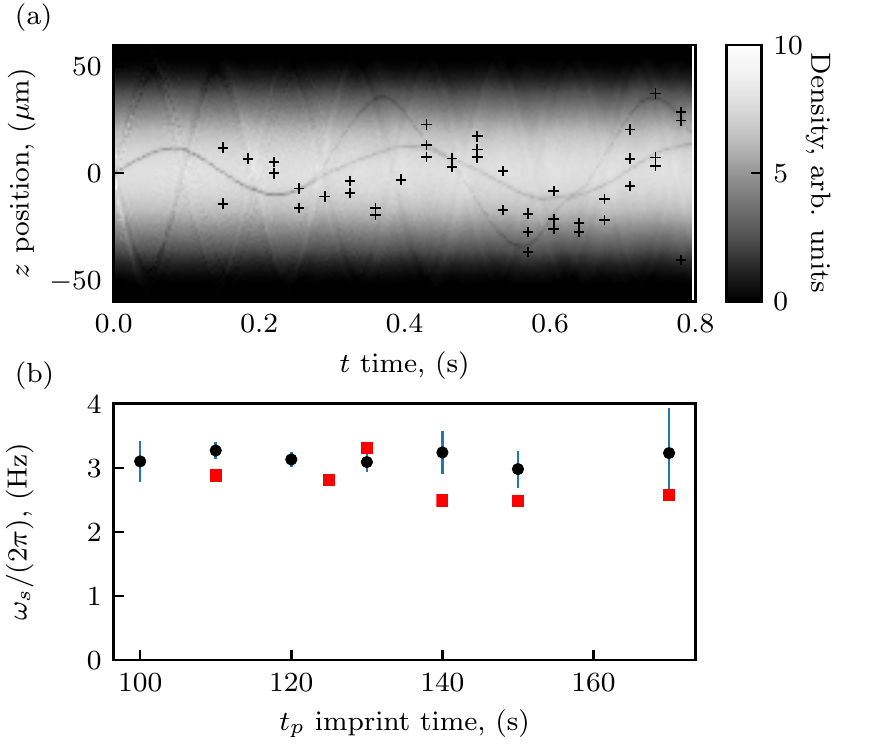}
     \caption{ Oscillation of the solitonic vortices (a) and oscillation frequencies (b).
    The black symbols on the grey colored integrated GPE wavefunction density in $\ez - t$ plane of (a) represent experimental data.
    Every set of experimental parameters was repeated three times, and a symbol is present for each clearly identified solitonic excitation. In cases in which more than one solitonic excitation was found in a single image we chose the position of the deepest depletion. The simulations use $t_p=140\ \mu\mathrm{s}$. 
     The symbols, red squares and black circles, in (b) represent GPE and experimental calculations, respectively, where the error bar represents the single-$\sigma$ uncertainty  obtained from the sinusoidal fits.
     }
    \label{fig:PHE1D}
\end{figure}

Figure~\ref{fig:PHE2D0} shows the 3d density projected onto the $\ex-\ez$ and $\ey-\ez$ planes, i.e., $|\psi(x,z)|^2$ and $|\psi(y,z)|^2$, and the associated planar phase cuts, immediately following the application of $V_\text{step}$ for a phase imprinting time $t_p = 140\ \mu\text{s}$. 
The phase-imprinted planar dark soliton subsequently evolves into a vortex ring as shown in ~\cite{supple}.
%. This is shown in ~\cite{supple} by a vortex pair in both the $\ex-\ez$ and $\ey-\ez$ planes.
This ring eventually decays~\cite{PhysRevA.98.033609} into a pair of solitonic vortices aligned along the direction of stronger transverse confinement and exhibits a large amplitude oscillation ~\cite{supple}. 
The weak interaction of the resulting
structures with the background of excited phonons leads to effective (apparent) dissipative longitudinal dynamics and to the selection of two separate trajectories, one for each solitonic vortex as shown in Figs.~\ref{fig:PHE2D} and~\ref{fig:PHE1D}. 
%Additionally, it is causing the oscillation amplitude to increase, approaching the longitudinal Thomas-Fermi (TF) radius at larger times as shown in the top panel of Fig.~\ref{fig:PHE1D}. 
{In the experimental results, over long times, the amplitude
of oscillation appears to be growing, possibly reflecting the (weak)
anti-damping effect due to the thermal fraction~\cite{prouka}.}
%\ibs{I associate this with anti-damping from scattering off of the thermal atoms, but with nomically constant thermal fraction.}
Both the numerical and experimental results (black crosses) find an oscillation frequency $\omega_s$
 %\approx 2\pi \times 3\ \text{Hz}$
far from the ``canonical'' result of $\omega_s =  \omega_z / \sqrt{2}$ for planar solitons~\cite{PhysRevLett.84.2298,Frantzeskakis_2010}. 
 %This gives $w_s\sim w_z/\sqrt{9}$, the expected frequency of oscillation of solitonic vortices 
Instead, the solitonic vortex oscillation frequency is 
significantly reduced~\cite{mateo2015stability,Fritsch2020creating}, e.g.,
 by a factor of about 3 in comparison to $\omega_z$, as shown in the bottom panel of Fig.~\ref{fig:PHE1D}. More generally, in line with~\cite{mateo2015stability}, we find that the relevant reduction is a function of the chemical potential, and the general agreement between the experimental data and the numerical simulations supports 
 solitonic vortex pair formation from the method of phase imprinting.
  %the rapid decay of a planar dark soliton 
%into a solitonic vortex pair.
This experiment-theory correspondence prompts us to further explore the stability of planar solitons and their transformation into other structures.

\section{Stability of planar dark solitons}

In this section we study the stability of dark solitary waves using a dimensionless version of Eq.~\eqref{eq:3DGP}
\begin{equation}
  i\frac{\partial\psi'}{\partial t'}= \left[-\frac{1}{2}(\nabla')^2+V'(\mathbf{r}')+g'|\psi'|^2 \right] \psi',
  \label{eq:3DGPd}
\end{equation}
in terms of the dimensionless quantities $t'=\omega_{\perp} t$, $\mathbf{r}'=\mathbf{r}/a_{\perp}$, $\psi'=a_{\perp}^{3/2} \psi$, $V(\mathbf{r}')=[l_x^2 (x')^2+l_y^2 (y')^2+l_z^2 (z')^2]/2$, $g'=4 \pi a_s/a_{\perp}$, and $\mu'=\mu/(\hbar \omega_{\perp})$.
Here, we use the transverse degrees of freedom to define the natural units of frequency $\omega_{\perp}=\sqrt{\omega_x \omega_y}$ and length $a_{\perp}=\sqrt{\hbar/(m \omega_{\perp})}$. 
In addition, we introduce anisotropy parameters $l_x=\omega_x/\omega_{\perp}$, $l_y=\omega_y/\omega_{\perp}$ and $l_z=\omega_z/\omega_{\perp}$.

In line with the last section, we use a chemical potential $\mu= h\times 1.1\ \mathrm{kHz}$, longitudinal frequency $\omega_z/(2\pi)=10\ \mathrm{Hz}$, and transverse frequency ranging from $\omega_{\perp}/(2\pi)=100\ \mathrm{Hz}$ to $1\ \mathrm{kHz}$. In what follows, we take $\omega_x=\omega_y$ for simplicity.

The 1D counterpart of Eq.~\eqref{eq:3DGP} 
with $V=0$ supports the dark soliton solution~\cite{kevrekidis2015defocusing}
\begin{equation}
 \psi_k'(z',t')=\sqrt{n_0} \left[B \tanh(\zeta)+i A\right]e^{i(kz'-\omega t')}.
  \label{eq:1D_dark}
\end{equation}
Here, $n_0=\mu'$, $\zeta=\sqrt{n_0}B[z'-z_0(t')]$, $z_0=vt'+Z_0$, $v=A\sqrt{n_0}+k$, $\omega=k^2/2+n_0$ and $A^2+B^2=1$, and restrict ourselves to $k=0$ solutions (of a stationary background). The remaining properties have the standard meaning,
i.e., $v$ is the soliton speed, $n_0$ the background density,
$z_0$ marks the center of mass (and $Z_0$ is its initial value), 
while $B$ is associated with the
inverse solitonic width~\cite{kevrekidis2015defocusing}.
%For non-zero wavenumber $k$ this corresponds to a solution moving with velocity $v$ with respect to the background BEC.

\subsection{Stationary solutions}
%\af{Ian can correct me if I'm wrong, but I think it is standard to use roman letters with abbreviations, like  $\psi_{\mathrm{ds}}$, $\lambda_{\mathrm{r}}$, $\lambda_{\mathrm{i}}$, $c_{\mathrm{s}}$}

In the presence of a trap, the wavefunction for a 3d stationary solution is approximately  
\begin{equation}
  \psi'(\mathbf{r'},t')=\psi_{ds}(z')\Phi(\mathbf{r'})e^{-i \mu' t'},
  \label{eq:1D_darks}
\end{equation}
  where $\Phi(\mathbf{r'})$ represents the ground state Thomas-Fermi (TF) wavefunction~\cite{pethick,stringari} and 
  $\psi_{ds}$ represents the stationary dark soliton from
  Eq.~(\ref{eq:1D_dark}), associated with $A=v=0$. 
\begin{figure*}
     \includegraphics{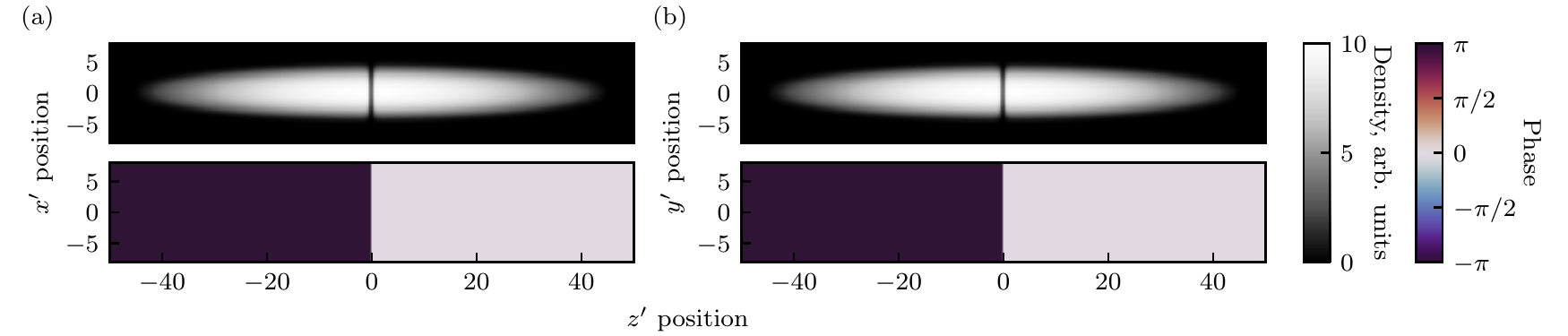}
       \caption{Condensate GPE density of the Eq.~\eqref{eq:ste} (top) and phase cross-section (bottom) show the stationary 3D dark soliton solution for $\omega_{\perp}/(2\pi)=100\ \mathrm{Hz}$, where $\omega_x=\omega_y$ and $v=0$. The lower panel shows the soliton phase jump from  $\pi$ (dark) to 0 (light). 
       (a) and (b) plot $|\psi'(x',0,z')|^2$ and $|\psi'(0,y',z')|^2$ respectively.
       }
    \label{fig:DSS}
\end{figure*}
Using Eq.~\eqref{eq:1D_darks} as an initial guess for a root finding algorithm, we obtain numerically exact solutions by solving
\begin{equation}
   \mathcal{F}(\psi') = \left[-\frac{1}{2}\nabla^2+V(\mathbf{r'})+g'|\psi'|^2 -\mu' \right]\psi'=0.
      \label{eq:ste}
  \end{equation}
Fig.~\ref{fig:DSS} shows the solution of Eq.~\eqref{eq:ste} for $\omega_{\perp}/(2\pi)=100$~Hz. We now 
%deal with the problem associated with the identification of the 
explore both the stable and unstable modes of $\psi'$ by identifying the eigenvalues associated with the BdG equations~\cite{pethick,stringari,kevrekidis2015defocusing} 
for the parameter range of $\omega_{\perp}$.
For relevant details, see the Appendix \ref{app:ES}. An eigenvalue $\lambda$ has a  real part denoted by $\lambda_r$ and an imaginary part denoted by $\lambda_i$. It is important to recall that a non-vanishing $\lambda_r$, given the  Hamiltonian nature of our GPE (\ref{eq:3DGPd}), is tantamount to the presence of a dynamical instability.
   \begin{figure}
 \includegraphics[width=\columnwidth,angle =0]{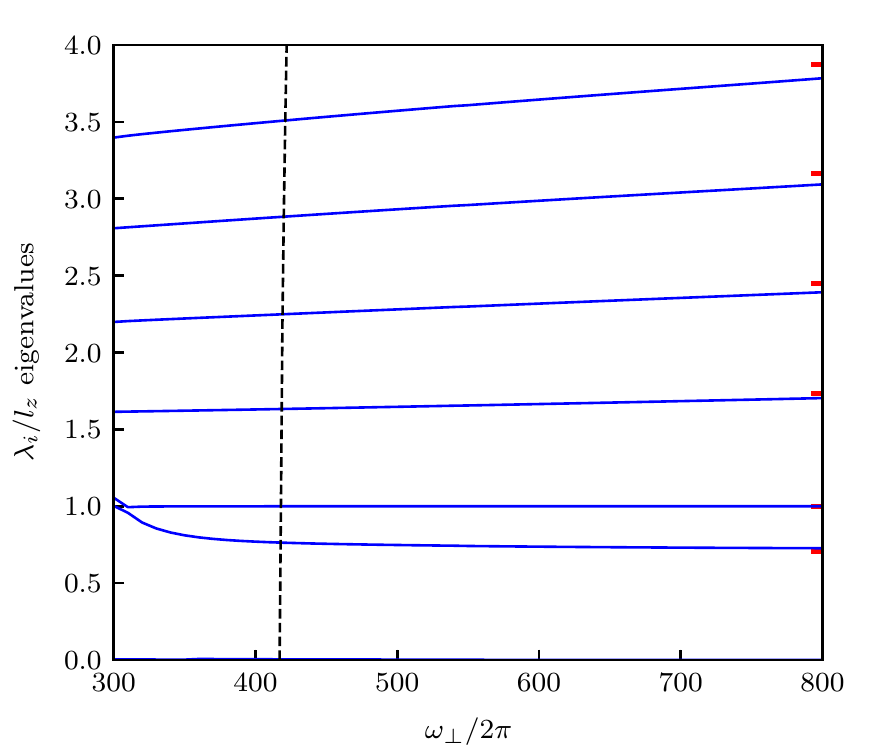}
       \caption{Imaginary eigenvalues $\lambda_i$ obtained from the BdG analysis  plotted as a function of $\omega_{\perp}$ for $\omega_x=\omega_y$. 
       The vertical dashed black line represents the mode responsible for instability of the planar
dark soliton, while blue lines represent stable modes. In the limit of large $\omega_{\perp}$, the 1d analytically available results, indicated by the red short lines, are retrieved (see text for further details).
    %   \pgk{Mithun: you need to explain the vertical dashed   line. }
      % \ibs{Add text labels next to the modes on the right indicating each quantum number.  Make the ``vertical'' mode a different color since it is special.} 
      }
    \label{fig:DSEigim}
\end{figure}
   \begin{figure}
 \includegraphics[width=\columnwidth,angle =0]{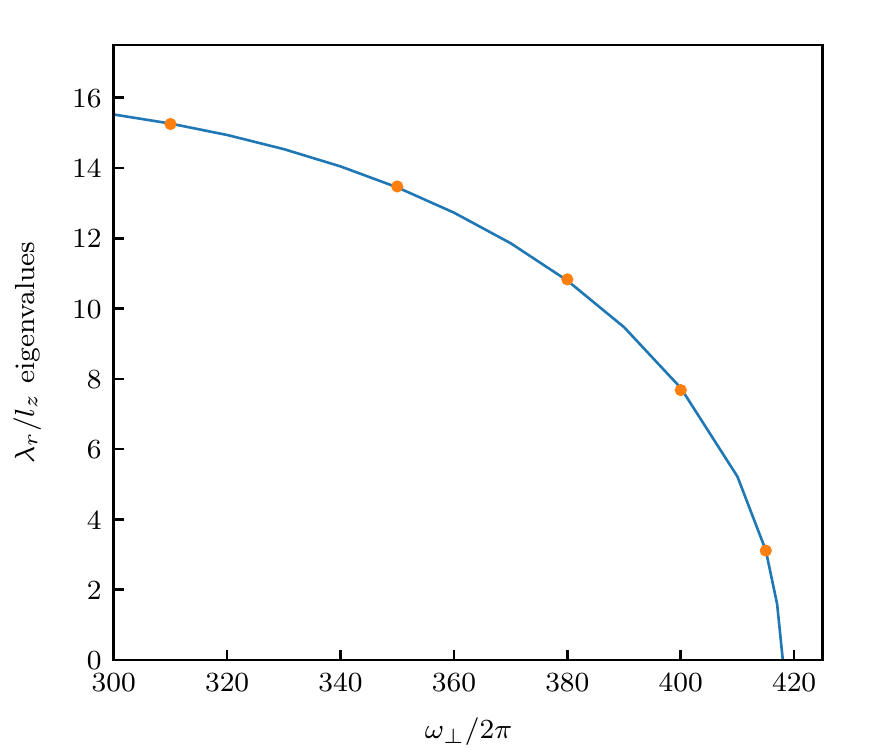}
       \caption{The largest real eigenvalue $\lambda_r$, 
       associated with the
       instability of stationary, planar dark soliton, as a function of $\omega_{\perp}$ for $\omega_x=\omega_y$. The red circles are obtained from GPE dynamics and blue line represents BdG analysis. 
       %The inset shows $\lambda^2$ as a function of  $\omega_{\perp}$. 
      % \pgk{Mithun, Ian is right that given the scale of the graph        the inset is not particularly meaningful...How about you removeit??}
      % \ibs{In the inset the quantity seems to go negative which makes no sense for a squared quantity.  Also why is it being plotted squared?  What does the inset add?}
       }
    \label{fig:DSEigr}
\end{figure}

There exist two limiting cases for the stability problem of the dark solitary waves in a trapped condensate, more
specifically, the TF limit at large chemical
potential $\mu'$ and the linear limit at small $\mu'$. 
As is well-known~\cite{kevrekidis2015defocusing} 
in the 1d limit of the GPE model, the anomalous mode ($\omega_0$), corresponding to negative energy is associated with the oscillation of the single dark soliton with a frequency of $\approx l_z /\sqrt{2}$~\cite{PhysRevLett.84.2298}.
%\cite{kevrekidis2017stability}. 
The other modes $\omega_n$ (associated with the
vibration modes of the background cloud) are expected to approach the frequencies of $l_z\sqrt{n(n+1)/2}$~\cite{stringari}. On the other hand, in the linear limit of vanishing density $|\psi'|^2 \rightarrow 0$, the eigenvalue problem of Eq.~\eqref{eq:3DGP} reduces to that of a {\it linear} quantum harmonic oscillator with corresponding energies of eigenmodes $|k,l,m\rangle$ 
\begin{align}
E_{k,l,m} &= \left(k+ \frac{1}{2}\right) l_x+ \left(l+ \frac{1}{2}\right) l_y+
\left(m + \frac{1}{2}\right) l_z.
\end{align}
%\ibs{please update to have k,l,m.  Has a zero point energy been removed?}
A single dark soliton at $z'=0$ (pertaining to the $|0,0,1\rangle$ in the linear limit) has energy $E_{0,0,1} = 
(l_x + l_y+ 3 l_z)/2$.
To explore the validity of these predictions, we rescale the imaginary eigenvalues by $l_z$ and show the corresponding results in Fig.~\ref{fig:DSEigim}. 
Indeed, our numerically calculated imaginary eigenvalues confirm the expectations on the basis of the 1d
predictions, as our numerical continuation results
approach progressively the 1d TF limit as $\omega_{\perp}$
increases. We note that the chemical potential controls the approach to the TF limit for the present setting. The mode responsible for the instability of the planar
dark soliton appears as a nearly vertical dashed black line in 
Fig.~\ref{fig:DSEigim}.
%and we thus now analyze the relevant mode in further detail.

%
   \begin{figure*}
 \includegraphics[width=\textwidth,angle =0]{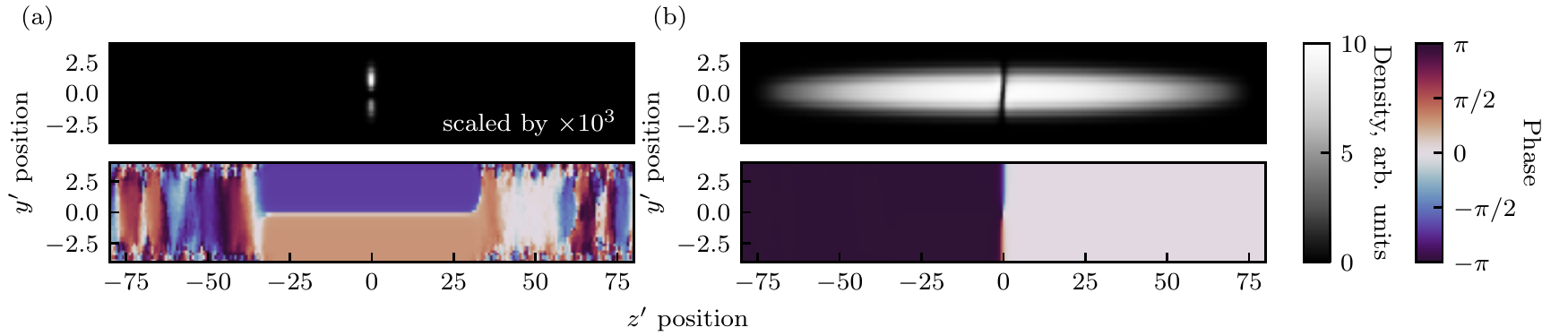}
       \caption{(left)  The eigenvector $|\varphi|^2$ corresponds to the largest real eigenvalue of Eq.~\eqref{eq:gpeeig} and its phase profile for the parameter $\omega_{\perp}/(2\pi)=415$~Hz and $\omega_x=\omega_y$. (right panel) The density $|\psi'+\epsilon \varphi|^2$, where $\psi'$ is the steady state wavefunction and $\epsilon=20$. Lower panels in (a) and (b) show the phase cross-section.
      % \ibs{In this and all of the graphs with the density color bars in "arb units" simply make them run from 0 to 10. In this graph remove the color bar on the left and add white text inside the black background saying ``scaled by $\times 10^3$'' or whatever is needed to make correctly scaled compared to the color bar in (b). }
      }
    \label{fig:DSEigv415}
\end{figure*}
The analysis of real eigenvalues displayed in Fig.~\ref{fig:DSEigr} shows that, as $\omega_{\perp}$  decreases (going away from the above mentioned quasi-1d limit), there exists an instability window for smaller values of the relevant parameter. The critical $\omega_{\perp}$, namely $\omega_{\perp}^c$, at which $\lambda_r$ crosses zero is at $2\pi\times 418\ \mathrm{Hz}$. Figure~\ref{fig:DSEigv415} shows the eigenvector corresponding to the largest real eigenvalue of Eq.~\eqref{eq:gpeeig} for the parameter $\omega_{\perp}/(2\pi)=415\ \mathrm{Hz}$ and $\omega_x=\omega_y$. Indeed, on the basis of the relevant unstable eigenmode, we expect the planar dark solitary wave structure to break its axial symmetry and reshape itself in the form of a vortex ring as a result of the relevant growing mode, recalling that the latter only breaks up towards a pair of solitonic vortices considerably later in the dynamical evolution. It is interesting to highlight that the above instability manifestation
appears to
be a natural 3d generalization of the destabilization
of a 2d planar dark soliton towards the formation of
a vortex dipole, explored from a bifurcation 
theory perspective, e.g., in~\cite{PhysRevA.77.053610,PhysRevA.82.013646}.

\begin{figure*}
 \includegraphics[width=\textwidth,angle =0]{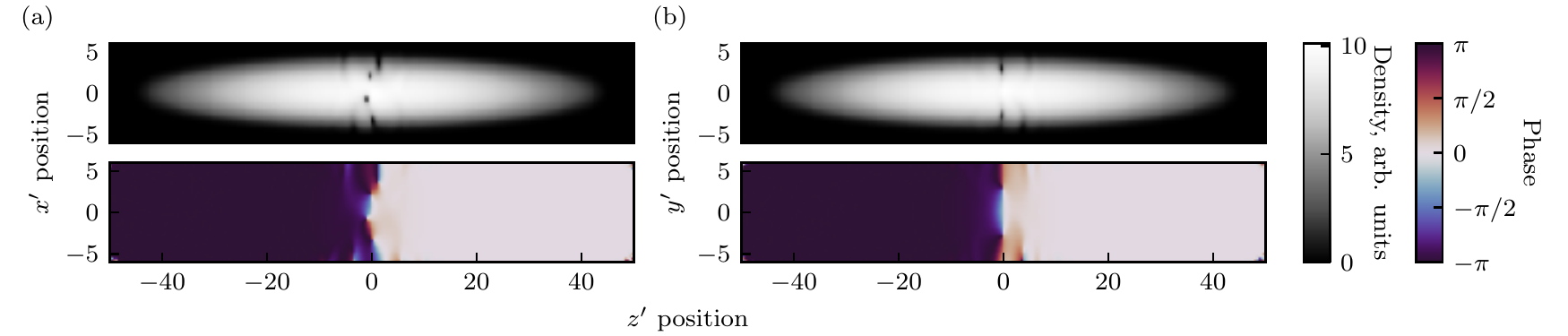}
       \caption{Condensate densities (top) and phase cross-section (bottom) show the existence of a vortex ring at $t'=15$ after evolving the steady state dark soliton solution. Here, $\omega_{\perp}/(2\pi)=100\ \mathrm{Hz}$ and $\omega_x=\omega_y$. (a) and (b) plot $|\psi(x',0,z')|^2$ and $|\psi(0,y',z')|^2$ respectively.} 
    \label{fig:DSE}
\end{figure*}

\begin{figure*}
 \includegraphics[width=\textwidth,angle =0]{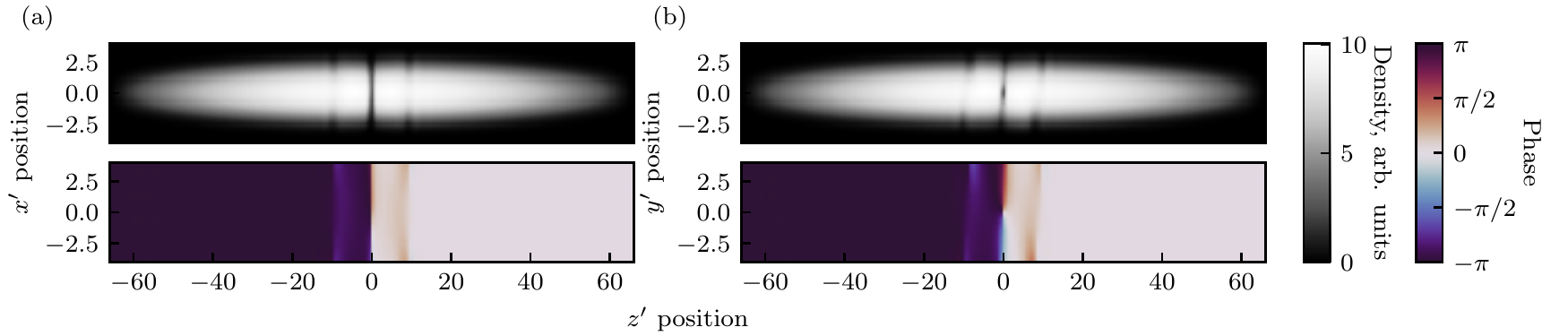}
       \caption{Condensate densities (top) and phase cross-section (bottom) show a solitonic vortex at $t'=50$ after evolving the steady state dark soliton solution. Here, $\omega_{\perp}/(2\pi)=250\ \mathrm{Hz}$ and $\omega_x=\omega_y$. (a) and (b) plot $|\psi'(x',0,z')|^2$ and $|\psi'(0,y',z')|^2$ respectively.
       %\ibs{why is this labeled $x$ and etc while Fig. 7,9, and 11 are primed?}
       }
    \label{fig:DSVR}
\end{figure*}

\begin{figure}
 \includegraphics[width=\columnwidth,angle =0]{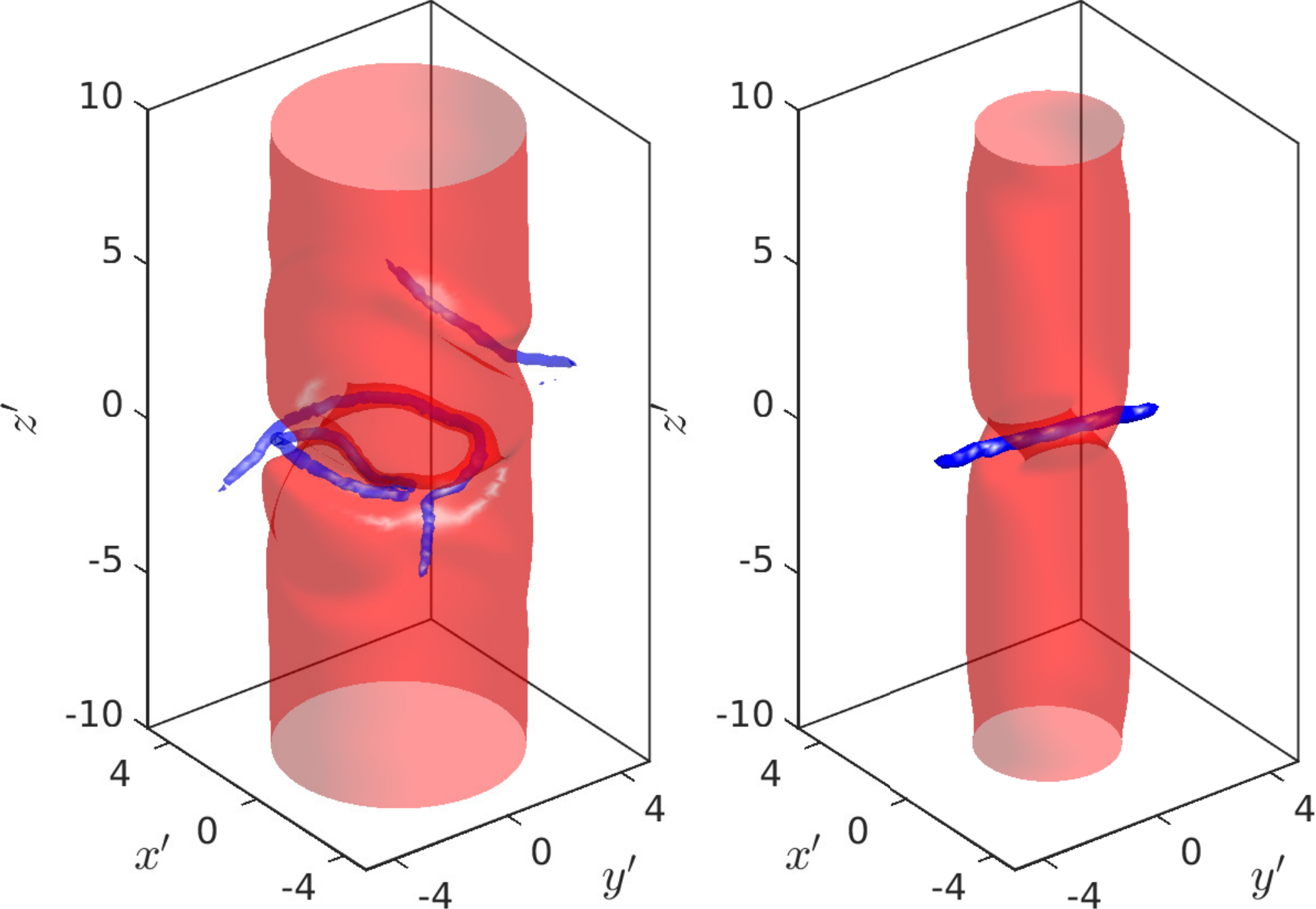}
       \caption{The three dimensional view of the snapshots shown in Fig.~\ref{fig:DSE} (left) and Fig.~\ref{fig:DSVR} (right). The red and blue colors represent cuts of constant density and vorticity. %  \ibs{These are actually plotted with a left-handed coordinate system!}
     }
    \label{fig:DS3Dview}
\end{figure}

The most unstable eigenvalue (the one bearing the largest real part) can be computed from the GPE dynamics starting from the perturbed steady state solutions of Eq.~\eqref{eq:ste}~\cite{PhysRevA.105.053303}.
We evolve such waveforms starting with a small perturbation $\delta \psi$ 
[$\approx 10^{-10}$ relative to the $\mathrm{max}(|\psi'|)$].
As expected from the linear
stability analysis, we find that this perturbation grows exponentially as $\psi'=\psi_0'\exp(\lambda_r t')$ with an instability growth time $\tau=-\log(|\delta \psi|)/\lambda_r$.
The computed growth rate $\lambda_r$ via this ``direct integration'' approach,
indicated by circles on Fig.~\ref{fig:DSEigr}, matches well with the largest eigenvalues obtained from the spectral analysis of the BdG equations, thus confirming the robustness of our numerical findings.
The GPE dynamics for different $\omega_{\perp}$ shows that the strength of transverse confinement determines the resulting wave structures induced by the transverse instability of dark solitary waves.
We generally find that tighter confinement favors the formation of a solitonic vortex, while looser traps result in a long-lived vortex ring.
Figs.~\ref{fig:DSE} and \ref{fig:DSVR} show the vortex ring and solitonic vortex generated during the GPE dynamics with $\omega_{\perp}/(2\pi)=100\ \mathrm{Hz}$ and $\omega_{\perp}/(2\pi)=250\ \mathrm{Hz}$.
Their existence can be further confirmed from the 3D condensate densities shown in Fig.~\ref{fig:DS3Dview}.
%Though the system favors vortex ring formation for smaller transverse confinements, it further decay into solitonic vortex pairs at the later stage of evolution.
The single solitonic vortex formation from the imprinted dark soliton for higher $\omega_{\perp}$ is due to the asymmetric flow (see Fig.~\ref{fig:DSEigv415}) arising from the transverse instability that leads to the bending of density on the $\ey-\ez$ plane \cite{PhysRevLett.113.065302}.
%--------------------- %
%

\subsection{Traveling solutions}

   \begin{figure*}
     \includegraphics[width=\textwidth,angle =0]{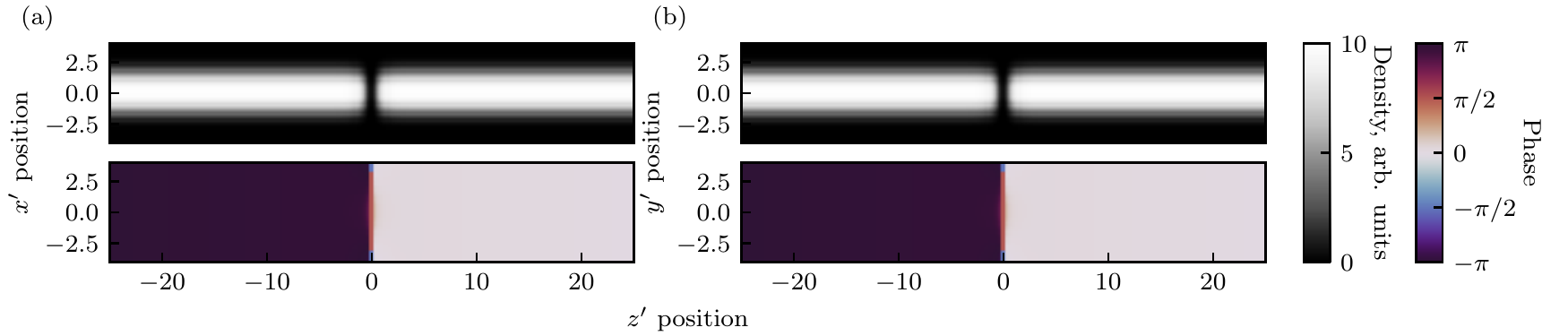}
       \caption{Condensate densities (top) and phase cross-section (bottom) of numerically exact traveling planar dark soliton solution for $\omega_{\perp}/(2\pi)=400$~Hz, where $\omega_x=\omega_y$ and $v=0.1c_s$. (a) and (b) plot $|\psi(x',0,z')|^2$ and $|\psi(0,y',z')|^2$ respectively.}
    \label{fig:vDSS}
\end{figure*}
   \begin{figure}
 \includegraphics[width=\columnwidth,angle =0]{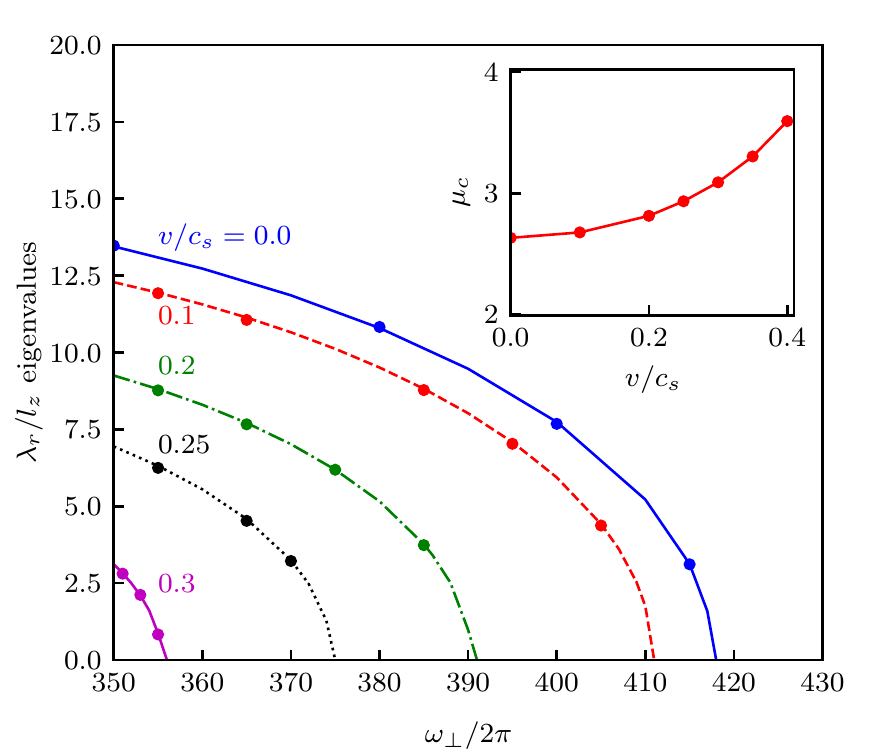}
       \caption{The largest real eigenvalues leading to the unstable dynamics of moving solitons as a function of $\omega_{\perp}$.The lines of the main figure (top to bottom) represent the cases, $v=(0,0.1,0.2,0.25,0.3)\times c_s$ obtained from the eigenvalue problem, while circles represent the growth rate obtained from the GPE dynamics. The inset shows the critical chemical potential $\mu_c$ as a function of $v/c_s$.
       }
    \label{fig:eigvelo}
\end{figure}

We now consider the case of traveling solitary waves. In this case, we find the steady solution of the 3D-GPE equation  
in the so-called co-traveling frame variant of
Eq.~\eqref{eq:ste} for which
\begin{align*}
 \mathcal{F}(\psi') &= \left[-\frac{1}{2}\nabla^2+iv\frac{\partial}{\partial z'}+V(\mathbf{r'})+g'|\psi'|^2 -\mu' \right] \psi' =0
%\label{TW}
\end{align*}
and $v$ represents the soliton velocity 
i.e., by seeking a numerically exact stationary solution
in the co-traveling frame along $\ez$, we obtain a
traveling solution of the original frame.
In this case, we fix the longitudinal length $L_z=120 a_{\perp}$ and remove the trap along the z-direction, so that
genuine traveling can be feasible along this direction. Fig.~\ref{fig:vDSS} shows the steady state solution for $v=0.1c_s$, where $c_s=\sqrt{\mu'}$ is the sound velocity, and $\omega_{\perp}/(2\pi)=400\ \mathrm{Hz}$.
This shows a phase jump which corresponds to the gray soliton~\cite{Frantzeskakis_2010} at the $\mathbf{r'}=0$ plane. 
The additional phase jump 
%$2\pi$ 
at the boundary is due to the implemented periodic boundary condition. 
%pgk{Mithun: at the boundary I am not sure that the phase jump is $2 \pi$.  It is  certainly not obvious in the
%phase plots. I am not sure how it is best to present
%this result? Perhaps in a slightly smaller domain
%and avoid this sentence altogether?}

\begin{figure}
 \includegraphics[width=0.8\columnwidth,angle =0]{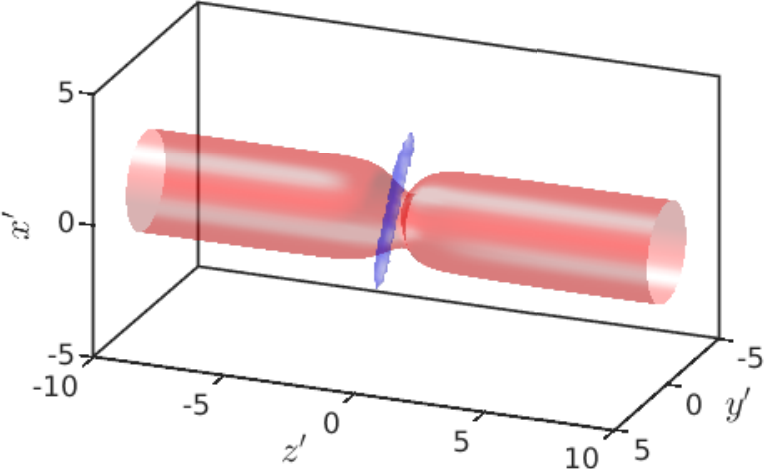}
       \caption{The three dimensional view of condensate density $|\psi'|^2$ at $t'=80$ for $v=0.2c_s$. The red and blue colors represent cuts of constant density and vorticity.
      % \ibs{Make the aspect ratio of this the same as in Fig. 10 so they are easy to compare.  Actually is there a reason not to make this part of Fig. 10?}
      }
    \label{fig:DS3Dviewv02}
\end{figure}

Importantly, the computation of the traveling 
planar dark solitary wave solution is accompanied by
the solution of the BdG equations to 
identify the spectral stability of the relevant
state. 
The largest real eigenvalues leading to the unstable dynamics of moving solitons are shown in Fig.~\ref{fig:eigvelo} as a function of $\omega_{\perp}$. As the velocity increases, the $\lambda_r/l_z$ decreases and, in this way, leads to the  increase in the critical transverse frequency $\omega_c$, which representing the transition point of stable to unstable dynamics. The corresponding chemical potential $\mu_c = \mu/(\hbar \omega_{\perp}^c)$ is shown in the inset of Fig.~\ref{fig:eigvelo} as a function of $v/c_s$. This corroborates, through detailed 3D computations, 
the earlier dynamical results of~\cite{PhysRevLett.89.110401}. On the other hand, we find from the GPE dynamics that the transverse instability leads to the formation of solitonic vortices as in the case of $v=0$. Fig.~\ref{fig:DS3Dviewv02} shows the solitonic vortex at $t'=80$ formed during the dynamical evolution of the condensate density for $v=0.2c_s$.

\section{Conclusions and outlook}

Motivated by experimental observations in 3D BECs indicative of the dynamical instability of planar dark solitons, we analyzed the stability of both stationary and traveling planar dark solitons.
We first illustrated the comparison of our 3D time-dependent numerical computations with the experimentally observed time-evolution. 
This indicated that, for the experimental parameters, the phase imprinting of a planar dark soliton gave rise to a vortex ring
%gave rise to a planar dark soliton 
that subsequently decayed into a solitonic vortex. 
%For weaker transverse confinement, it could instead decay into a long lived vortex ring. These findings 
This finding motivated us to explore the stability of planar dark solitons as a function of the transverse confinement, thereby quantifying the transition from the 3D to the 1D regime. 
Indeed, for sufficiently strong transverse confinement, 
our BdG analysis identified the quasi-1d limiting values of the corresponding excitation frequencies. This analysis further captured the destabilization of planar dark solitons in the 3D regime.
We identified the most rapidly growing unstable mode in two distinct ways{; one by the GPE direct simulation and another by the BdG analysis,} and monitored the resulting dynamics leading to the formation of solitonic vortices or vortex rings (depending on the transverse confinement).
 
Although this problem has been studied from a range of perspectives, there are still numerous open questions. 
Our primary focus here
is on the implications of the transverse confinement.
However, these structures are not exact solutions for non-zero longitudinal confinement, therefore their stability and properties as a function of longitudinal confinement 
merits further study. 
Moreover, a careful inspection of the movie in~\cite{supple} raises
intriguing questions on the scattering interactions between 
solitonic vortices in confined geometries.
In the same vein, the behavior of solitonic structures in box potentials~\cite{PhysRevLett.110.200406} is of considerable
current interest.
In this case, one can study how structures that are near-exact solutions in the bulk abruptly impinge on the system's edge.
Further possibilities could involve localized barriers~\cite{PhysRevA.94.063612} or even random potentials~\cite{doi:10.1073/pnas.1615004114} where the soliton's fate and trajectory are unknown.

\begin{acknowledgments}
We gratefully acknowledge G.~H.~Reid, and K.~Rodriguez for  meticulously reading the manuscript, R.~Carretero-Gonz\'{a}lez for fruitful discussions; and Mason Porter for long ago initiating our exchanges on this topic.
%P.G.K. was supported by the US National Science Foundation (NSF) under Grant No.~PHY-2110030.
This material is based upon work supported by the US
National Science Foundation under Grants No. PHY-2110030 and DMS-2204702 (P.G.K.).
I.B.S. and A.F. were partially supported by the National Institute of Standards and Technology, as well as the NSF through: the Physics Frontier Center at the Joint Quantum Institute (PHY-1430094) and the Quantum Leap Challenge Institute for Robust Quantum Simulation (OMA-2120757).
\end{acknowledgments}

%----------------------------------------
\appendix
\section{Linearized GP equation}\label{app:ES}
The adimensionalized GPE can be written as
\begin{equation}
   i\frac{\partial\psi'}{\partial t'}= \left[-\frac{1}{2}\nabla^2+V(\mathbf{r'})-\mu'+g'n \right] \psi,
\end{equation}
with $n=|\psi'|^2$.
The linearized evolution $\psi'=\psi_0'+\varepsilon\,\delta\psi$ about the steady state $\psi_0'$ is described by
\begin{equation}
  i\frac{\partial\delta \psi}{\partial t'}= K_R\, \delta\psi+ K_I\, \delta\psi^{\ast},
\end{equation}
with
\begin{align*}
  K_R &=-\frac{1}{2}\nabla^2+V(\mathbf{r'})-\mu'+2 g |\psi'|^2 & \text{and} && K_I &= g' \psi'^2.
\end{align*}
Finally, for the perturbation $\delta \psi = P e^{\lambda t'}+ Q^{\ast} e^{\lambda^{\ast} t'}$, the linearized problem 
\begin{align}
i \lambda \left(
\begin{array}{c}
P \\ Q
\end{array}
\right) &= 
\left(
\begin{array}{cc}
K_R & K_I \\
-K_I^* & -K_R^* 
\end{array}
\right)
\left(
\begin{array}{c}
P \\ Q
\end{array}
\right).
\label{eq:gpeeig}
\end{align}
takes on the familiar BdG structure.

  %%%%%%%%%%%%%%
\bibliographystyle{apsrev4}
\let\itshape\upshape
\normalem
\bibliography{main}

\providecommand{\noopsort}[1]{}\providecommand{\singleletter}[1]{#1}%
\begin{thebibliography}{42}%
\makeatletter
\providecommand \@ifxundefined [1]{%
 \@ifx{#1\undefined}
}%
\providecommand \@ifnum [1]{%
 \ifnum #1\expandafter \@firstoftwo
 \else \expandafter \@secondoftwo
 \fi
}%
\providecommand \@ifx [1]{%
 \ifx #1\expandafter \@firstoftwo
 \else \expandafter \@secondoftwo
 \fi
}%
\providecommand \natexlab [1]{#1}%
\providecommand \enquote  [1]{``#1''}%
\providecommand \bibnamefont  [1]{#1}%
\providecommand \bibfnamefont [1]{#1}%
\providecommand \citenamefont [1]{#1}%
\providecommand \href@noop [0]{\@secondoftwo}%
\providecommand \href [0]{\begingroup \@sanitize@url \@href}%
\providecommand \@href[1]{\@@startlink{#1}\@@href}%
\providecommand \@@href[1]{\endgroup#1\@@endlink}%
\providecommand \@sanitize@url [0]{\catcode `\\12\catcode `\$12\catcode
  `\&12\catcode `\#12\catcode `\^12\catcode `\_12\catcode `\%12\relax}%
\providecommand \@@startlink[1]{}%
\providecommand \@@endlink[0]{}%
\providecommand \url  [0]{\begingroup\@sanitize@url \@url }%
\providecommand \@url [1]{\endgroup\@href {#1}{\urlprefix }}%
\providecommand \urlprefix  [0]{URL }%
\providecommand \Eprint [0]{\href }%
\providecommand \doibase [0]{http://dx.doi.org/}%
\providecommand \selectlanguage [0]{\@gobble}%
\providecommand \bibinfo  [0]{\@secondoftwo}%
\providecommand \bibfield  [0]{\@secondoftwo}%
\providecommand \translation [1]{[#1]}%
\providecommand \BibitemOpen [0]{}%
\providecommand \bibitemStop [0]{}%
\providecommand \bibitemNoStop [0]{.\EOS\space}%
\providecommand \EOS [0]{\spacefactor3000\relax}%
\providecommand \BibitemShut  [1]{\csname bibitem#1\endcsname}%
\let\auto@bib@innerbib\@empty
%</preamble>
\bibitem [{\citenamefont {Kevrekidis}\ \emph {et~al.}(2015)\citenamefont
  {Kevrekidis}, \citenamefont {Frantzeskakis},\ and\ \citenamefont
  {Carretero-Gonz{\'a}lez}}]{kevrekidis2015defocusing}%
  \BibitemOpen
  \bibfield  {author} {\bibinfo {author} {\bibfnamefont {P.~G.}\ \bibnamefont
  {Kevrekidis}}, \bibinfo {author} {\bibfnamefont {D.~J.}\ \bibnamefont
  {Frantzeskakis}}, \ and\ \bibinfo {author} {\bibfnamefont {R.}~\bibnamefont
  {Carretero-Gonz{\'a}lez}},\ }\href@noop {} {\emph {\bibinfo {title} {The
  defocusing nonlinear Schr{\"o}dinger equation: from dark solitons to vortices
  and vortex rings}}}\ (\bibinfo  {publisher} {SIAM},\ \bibinfo {year}
  {2015})\BibitemShut {NoStop}%
\bibitem [{\citenamefont {Frantzeskakis}(2010)}]{Frantzeskakis_2010}%
  \BibitemOpen
  \bibfield  {author} {\bibinfo {author} {\bibfnamefont {D.~J.}\ \bibnamefont
  {Frantzeskakis}},\ }\bibfield  {title} {\enquote {\bibinfo {title} {Dark
  solitons in atomic {Bose}-{Einstein} condensates: from theory to
  experiments},}\ }\href {\doibase 10.1088/1751-8113/43/21/213001} {\bibfield
  {journal} {\bibinfo  {journal} {Journal of Physics A: Mathematical and
  Theoretical}\ }\textbf {\bibinfo {volume} {43}},\ \bibinfo {pages} {213001}
  (\bibinfo {year} {2010})}\BibitemShut {NoStop}%
\bibitem [{\citenamefont {Kivshar}\ and\ \citenamefont
  {Luther-Davies}(1998)}]{KIVSHAR199881}%
  \BibitemOpen
  \bibfield  {author} {\bibinfo {author} {\bibfnamefont {Y.~S.}\ \bibnamefont
  {Kivshar}}\ and\ \bibinfo {author} {\bibfnamefont {B.}~\bibnamefont
  {Luther-Davies}},\ }\bibfield  {title} {\enquote {\bibinfo {title} {Dark
  optical solitons: physics and applications},}\ }\href {\doibase
  https://doi.org/10.1016/S0370-1573(97)00073-2} {\bibfield  {journal}
  {\bibinfo  {journal} {Physics Reports}\ }\textbf {\bibinfo {volume} {298}},\
  \bibinfo {pages} {81} (\bibinfo {year} {1998})}\BibitemShut {NoStop}%
\bibitem [{\citenamefont {Denardo}\ \emph {et~al.}(1992)\citenamefont
  {Denardo}, \citenamefont {Galvin}, \citenamefont {Greenfield}, \citenamefont
  {Larraza}, \citenamefont {Putterman},\ and\ \citenamefont
  {Wright}}]{PhysRevLett.68.1730}%
  \BibitemOpen
  \bibfield  {author} {\bibinfo {author} {\bibfnamefont {B.}~\bibnamefont
  {Denardo}}, \bibinfo {author} {\bibfnamefont {B.}~\bibnamefont {Galvin}},
  \bibinfo {author} {\bibfnamefont {A.}~\bibnamefont {Greenfield}}, \bibinfo
  {author} {\bibfnamefont {A.}~\bibnamefont {Larraza}}, \bibinfo {author}
  {\bibfnamefont {S.}~\bibnamefont {Putterman}}, \ and\ \bibinfo {author}
  {\bibfnamefont {W.}~\bibnamefont {Wright}},\ }\bibfield  {title} {\enquote
  {\bibinfo {title} {Observations of localized structures in nonlinear
  lattices: Domain walls and kinks},}\ }\href {\doibase
  10.1103/PhysRevLett.68.1730} {\bibfield  {journal} {\bibinfo  {journal}
  {Phys. Rev. Lett.}\ }\textbf {\bibinfo {volume} {68}},\ \bibinfo {pages}
  {1730} (\bibinfo {year} {1992})}\BibitemShut {NoStop}%
\bibitem [{\citenamefont {Marquie}\ \emph {et~al.}(1994)\citenamefont
  {Marquie}, \citenamefont {Bilbault},\ and\ \citenamefont
  {Remoissenet}}]{PhysRevE.49.828}%
  \BibitemOpen
  \bibfield  {author} {\bibinfo {author} {\bibfnamefont {P.}~\bibnamefont
  {Marquie}}, \bibinfo {author} {\bibfnamefont {J.~M.}\ \bibnamefont
  {Bilbault}}, \ and\ \bibinfo {author} {\bibfnamefont {M.}~\bibnamefont
  {Remoissenet}},\ }\bibfield  {title} {\enquote {\bibinfo {title} {Generation
  of envelope and hole solitons in an experimental transmission line},}\ }\href
  {\doibase 10.1103/PhysRevE.49.828} {\bibfield  {journal} {\bibinfo  {journal}
  {Phys. Rev. E}\ }\textbf {\bibinfo {volume} {49}},\ \bibinfo {pages} {828}
  (\bibinfo {year} {1994})}\BibitemShut {NoStop}%
\bibitem [{\citenamefont {Chen}\ \emph {et~al.}(1993)\citenamefont {Chen},
  \citenamefont {Tsankov}, \citenamefont {Nash},\ and\ \citenamefont
  {Patton}}]{PhysRevLett.70.1707}%
  \BibitemOpen
  \bibfield  {author} {\bibinfo {author} {\bibfnamefont {M.}~\bibnamefont
  {Chen}}, \bibinfo {author} {\bibfnamefont {M.~A.}\ \bibnamefont {Tsankov}},
  \bibinfo {author} {\bibfnamefont {J.~M.}\ \bibnamefont {Nash}}, \ and\
  \bibinfo {author} {\bibfnamefont {C.~E.}\ \bibnamefont {Patton}},\ }\bibfield
   {title} {\enquote {\bibinfo {title} {Microwave magnetic-envelope dark
  solitons in yttrium iron garnet thin films},}\ }\href {\doibase
  10.1103/PhysRevLett.70.1707} {\bibfield  {journal} {\bibinfo  {journal}
  {Phys. Rev. Lett.}\ }\textbf {\bibinfo {volume} {70}},\ \bibinfo {pages}
  {1707} (\bibinfo {year} {1993})}\BibitemShut {NoStop}%
\bibitem [{\citenamefont {Kalinikos}\ \emph {et~al.}(2000)\citenamefont
  {Kalinikos}, \citenamefont {Scott},\ and\ \citenamefont
  {Patton}}]{PhysRevLett.84.4697}%
  \BibitemOpen
  \bibfield  {author} {\bibinfo {author} {\bibfnamefont {B.~A.}\ \bibnamefont
  {Kalinikos}}, \bibinfo {author} {\bibfnamefont {M.~M.}\ \bibnamefont
  {Scott}}, \ and\ \bibinfo {author} {\bibfnamefont {C.~E.}\ \bibnamefont
  {Patton}},\ }\bibfield  {title} {\enquote {\bibinfo {title} {Self-generation
  of fundamental dark solitons in magnetic films},}\ }\href {\doibase
  10.1103/PhysRevLett.84.4697} {\bibfield  {journal} {\bibinfo  {journal}
  {Phys. Rev. Lett.}\ }\textbf {\bibinfo {volume} {84}},\ \bibinfo {pages}
  {4697} (\bibinfo {year} {2000})}\BibitemShut {NoStop}%
\bibitem [{\citenamefont {Achilleos}\ \emph {et~al.}(2015)\citenamefont
  {Achilleos}, \citenamefont {Richoux}, \citenamefont {Theocharis},\ and\
  \citenamefont {Frantzeskakis}}]{PhysRevE.91.023204}%
  \BibitemOpen
  \bibfield  {author} {\bibinfo {author} {\bibfnamefont {V.}~\bibnamefont
  {Achilleos}}, \bibinfo {author} {\bibfnamefont {O.}~\bibnamefont {Richoux}},
  \bibinfo {author} {\bibfnamefont {G.}~\bibnamefont {Theocharis}}, \ and\
  \bibinfo {author} {\bibfnamefont {D.~J.}\ \bibnamefont {Frantzeskakis}},\
  }\bibfield  {title} {\enquote {\bibinfo {title} {Acoustic solitons in
  waveguides with helmholtz resonators: Transmission line approach},}\ }\href
  {\doibase 10.1103/PhysRevE.91.023204} {\bibfield  {journal} {\bibinfo
  {journal} {Phys. Rev. E}\ }\textbf {\bibinfo {volume} {91}},\ \bibinfo
  {pages} {023204} (\bibinfo {year} {2015})}\BibitemShut {NoStop}%
\bibitem [{\citenamefont {Chabchoub}\ \emph {et~al.}(2013)\citenamefont
  {Chabchoub}, \citenamefont {Kimmoun}, \citenamefont {Branger}, \citenamefont
  {Hoffmann}, \citenamefont {Proment}, \citenamefont {Onorato},\ and\
  \citenamefont {Akhmediev}}]{PhysRevLett.110.124101}%
  \BibitemOpen
  \bibfield  {author} {\bibinfo {author} {\bibfnamefont {A.}~\bibnamefont
  {Chabchoub}}, \bibinfo {author} {\bibfnamefont {O.}~\bibnamefont {Kimmoun}},
  \bibinfo {author} {\bibfnamefont {H.}~\bibnamefont {Branger}}, \bibinfo
  {author} {\bibfnamefont {N.}~\bibnamefont {Hoffmann}}, \bibinfo {author}
  {\bibfnamefont {D.}~\bibnamefont {Proment}}, \bibinfo {author} {\bibfnamefont
  {M.}~\bibnamefont {Onorato}}, \ and\ \bibinfo {author} {\bibfnamefont
  {N.}~\bibnamefont {Akhmediev}},\ }\bibfield  {title} {\enquote {\bibinfo
  {title} {Experimental observation of dark solitons on the surface of
  water},}\ }\href {\doibase 10.1103/PhysRevLett.110.124101} {\bibfield
  {journal} {\bibinfo  {journal} {Phys. Rev. Lett.}\ }\textbf {\bibinfo
  {volume} {110}},\ \bibinfo {pages} {124101} (\bibinfo {year}
  {2013})}\BibitemShut {NoStop}%
\bibitem [{\citenamefont {Piccardi}\ \emph {et~al.}(2011)\citenamefont
  {Piccardi}, \citenamefont {Alberucci}, \citenamefont {Tabiryan},\ and\
  \citenamefont {Assanto}}]{Piccardi:11}%
  \BibitemOpen
  \bibfield  {author} {\bibinfo {author} {\bibfnamefont {A.}~\bibnamefont
  {Piccardi}}, \bibinfo {author} {\bibfnamefont {A.}~\bibnamefont {Alberucci}},
  \bibinfo {author} {\bibfnamefont {N.}~\bibnamefont {Tabiryan}}, \ and\
  \bibinfo {author} {\bibfnamefont {G.}~\bibnamefont {Assanto}},\ }\bibfield
  {title} {\enquote {\bibinfo {title} {Dark nematicons},}\ }\href {\doibase
  10.1364/OL.36.001356} {\bibfield  {journal} {\bibinfo  {journal} {Opt.
  Lett.}\ }\textbf {\bibinfo {volume} {36}},\ \bibinfo {pages} {1356} (\bibinfo
  {year} {2011})}\BibitemShut {NoStop}%
\bibitem [{\citenamefont {Carusotto}\ and\ \citenamefont
  {Ciuti}(2013)}]{carusotto_2013}%
  \BibitemOpen
  \bibfield  {author} {\bibinfo {author} {\bibfnamefont {I.}~\bibnamefont
  {Carusotto}}\ and\ \bibinfo {author} {\bibfnamefont {C.}~\bibnamefont
  {Ciuti}},\ }\bibfield  {title} {\enquote {\bibinfo {title} {Quantum fluids of
  light},}\ }\href {\doibase 10.1103/RevModPhys.85.299} {\bibfield  {journal}
  {\bibinfo  {journal} {Rev. Mod. Phys.}\ }\textbf {\bibinfo {volume} {85}},\
  \bibinfo {pages} {299} (\bibinfo {year} {2013})}\BibitemShut {NoStop}%
\bibitem [{\citenamefont {Pethick}\ and\ \citenamefont
  {Smith}(2008)}]{pethick}%
  \BibitemOpen
  \bibfield  {author} {\bibinfo {author} {\bibfnamefont {C.~J.}\ \bibnamefont
  {Pethick}}\ and\ \bibinfo {author} {\bibfnamefont {H.}~\bibnamefont
  {Smith}},\ }\href@noop {} {\emph {\bibinfo {title} {{Bose}-{Einstein}
  Condensation in Dilute Gases}}}\ (\bibinfo  {publisher} {Cambridge University
  Press},\ \bibinfo {address} {Cambridge, United Kingdom},\ \bibinfo {year}
  {2008})\BibitemShut {NoStop}%
\bibitem [{\citenamefont {Pitaevskii}\ and\ \citenamefont
  {Stringari}(2018)}]{stringari}%
  \BibitemOpen
  \bibfield  {author} {\bibinfo {author} {\bibfnamefont {L.}~\bibnamefont
  {Pitaevskii}}\ and\ \bibinfo {author} {\bibfnamefont {S.}~\bibnamefont
  {Stringari}},\ }\href@noop {} {\emph {\bibinfo {title} {{Bose}-{Einstein}
  Condensation and Superfluidity}}}\ (\bibinfo  {publisher} {Oxford University
  Press},\ \bibinfo {address} {Oxford, United Kingdom},\ \bibinfo {year}
  {2018})\BibitemShut {NoStop}%
\bibitem [{\citenamefont {Engels}\ and\ \citenamefont
  {Atherton}(2007)}]{PhysRevLett.99.160405}%
  \BibitemOpen
  \bibfield  {author} {\bibinfo {author} {\bibfnamefont {P.}~\bibnamefont
  {Engels}}\ and\ \bibinfo {author} {\bibfnamefont {C.}~\bibnamefont
  {Atherton}},\ }\bibfield  {title} {\enquote {\bibinfo {title} {Stationary and
  nonstationary fluid flow of a {Bose}-{Einstein} condensate through a
  penetrable barrier},}\ }\href {\doibase 10.1103/PhysRevLett.99.160405}
  {\bibfield  {journal} {\bibinfo  {journal} {Phys. Rev. Lett.}\ }\textbf
  {\bibinfo {volume} {99}},\ \bibinfo {pages} {160405} (\bibinfo {year}
  {2007})}\BibitemShut {NoStop}%
\bibitem [{\citenamefont {Becker}\ \emph {et~al.}(2008)\citenamefont {Becker},
  \citenamefont {Stellmer}, \citenamefont {{Soltan-Panahi}}, \citenamefont
  {D{\"o}rscher}, \citenamefont {Baumert}, \citenamefont {Richter},
  \citenamefont {Kronj{\"a}ger}, \citenamefont {Bongs},\ and\ \citenamefont
  {Sengstock}}]{Becker2008}%
  \BibitemOpen
  \bibfield  {author} {\bibinfo {author} {\bibfnamefont {C.}~\bibnamefont
  {Becker}}, \bibinfo {author} {\bibfnamefont {S.}~\bibnamefont {Stellmer}},
  \bibinfo {author} {\bibfnamefont {P.}~\bibnamefont {{Soltan-Panahi}}},
  \bibinfo {author} {\bibfnamefont {S.}~\bibnamefont {D{\"o}rscher}}, \bibinfo
  {author} {\bibfnamefont {M.}~\bibnamefont {Baumert}}, \bibinfo {author}
  {\bibfnamefont {E.-M.}\ \bibnamefont {Richter}}, \bibinfo {author}
  {\bibfnamefont {J.}~\bibnamefont {Kronj{\"a}ger}}, \bibinfo {author}
  {\bibfnamefont {K.}~\bibnamefont {Bongs}}, \ and\ \bibinfo {author}
  {\bibfnamefont {K.}~\bibnamefont {Sengstock}},\ }\bibfield  {title} {\enquote
  {\bibinfo {title} {Oscillations and interactions of dark and dark\textendash
  bright solitons in {{Bose}}\textendash{{Einstein}} condensates},}\ }\href
  {\doibase 10.1038/nphys962} {\bibfield  {journal} {\bibinfo  {journal} {Nat.
  Phys.}\ }\textbf {\bibinfo {volume} {4}},\ \bibinfo {pages} {496} (\bibinfo
  {year} {2008})}\BibitemShut {NoStop}%
\bibitem [{\citenamefont {Weller}\ \emph {et~al.}(2008)\citenamefont {Weller},
  \citenamefont {Ronzheimer}, \citenamefont {Gross}, \citenamefont {Esteve},
  \citenamefont {Oberthaler}, \citenamefont {Frantzeskakis}, \citenamefont
  {Theocharis},\ and\ \citenamefont {Kevrekidis}}]{PhysRevLett.101.130401}%
  \BibitemOpen
  \bibfield  {author} {\bibinfo {author} {\bibfnamefont {A.}~\bibnamefont
  {Weller}}, \bibinfo {author} {\bibfnamefont {J.~P.}\ \bibnamefont
  {Ronzheimer}}, \bibinfo {author} {\bibfnamefont {C.}~\bibnamefont {Gross}},
  \bibinfo {author} {\bibfnamefont {J.}~\bibnamefont {Esteve}}, \bibinfo
  {author} {\bibfnamefont {M.~K.}\ \bibnamefont {Oberthaler}}, \bibinfo
  {author} {\bibfnamefont {D.~J.}\ \bibnamefont {Frantzeskakis}}, \bibinfo
  {author} {\bibfnamefont {G.}~\bibnamefont {Theocharis}}, \ and\ \bibinfo
  {author} {\bibfnamefont {P.~G.}\ \bibnamefont {Kevrekidis}},\ }\bibfield
  {title} {\enquote {\bibinfo {title} {Experimental observation of oscillating
  and interacting matter wave dark solitons},}\ }\href {\doibase
  10.1103/PhysRevLett.101.130401} {\bibfield  {journal} {\bibinfo  {journal}
  {Phys. Rev. Lett.}\ }\textbf {\bibinfo {volume} {101}},\ \bibinfo {pages}
  {130401} (\bibinfo {year} {2008})}\BibitemShut {NoStop}%
\bibitem [{\citenamefont {Theocharis}\ \emph {et~al.}(2010)\citenamefont
  {Theocharis}, \citenamefont {Weller}, \citenamefont {Ronzheimer},
  \citenamefont {Gross}, \citenamefont {Oberthaler}, \citenamefont
  {Kevrekidis},\ and\ \citenamefont {Frantzeskakis}}]{PhysRevA.81.063604}%
  \BibitemOpen
  \bibfield  {author} {\bibinfo {author} {\bibfnamefont {G.}~\bibnamefont
  {Theocharis}}, \bibinfo {author} {\bibfnamefont {A.}~\bibnamefont {Weller}},
  \bibinfo {author} {\bibfnamefont {J.~P.}\ \bibnamefont {Ronzheimer}},
  \bibinfo {author} {\bibfnamefont {C.}~\bibnamefont {Gross}}, \bibinfo
  {author} {\bibfnamefont {M.~K.}\ \bibnamefont {Oberthaler}}, \bibinfo
  {author} {\bibfnamefont {P.~G.}\ \bibnamefont {Kevrekidis}}, \ and\ \bibinfo
  {author} {\bibfnamefont {D.~J.}\ \bibnamefont {Frantzeskakis}},\ }\bibfield
  {title} {\enquote {\bibinfo {title} {Multiple atomic dark solitons in
  cigar-shaped {Bose}-{Einstein} condensates},}\ }\href {\doibase
  10.1103/PhysRevA.81.063604} {\bibfield  {journal} {\bibinfo  {journal} {Phys.
  Rev. A}\ }\textbf {\bibinfo {volume} {81}},\ \bibinfo {pages} {063604}
  (\bibinfo {year} {2010})}\BibitemShut {NoStop}%
\bibitem [{\citenamefont {Fritsch}\ \emph {et~al.}(2020)\citenamefont
  {Fritsch}, \citenamefont {Lu}, \citenamefont {Reid}, \citenamefont
  {Pi\~neiro},\ and\ \citenamefont {Spielman}}]{Fritsch2020creating}%
  \BibitemOpen
  \bibfield  {author} {\bibinfo {author} {\bibfnamefont {A.~R.}\ \bibnamefont
  {Fritsch}}, \bibinfo {author} {\bibfnamefont {M.}~\bibnamefont {Lu}},
  \bibinfo {author} {\bibfnamefont {G.~H.}\ \bibnamefont {Reid}}, \bibinfo
  {author} {\bibfnamefont {A.~M.}\ \bibnamefont {Pi\~neiro}}, \ and\ \bibinfo
  {author} {\bibfnamefont {I.~B.}\ \bibnamefont {Spielman}},\ }\bibfield
  {title} {\enquote {\bibinfo {title} {Creating solitons with controllable and
  near-zero velocity in {Bose}-{Einstein} condensates},}\ }\href {\doibase
  10.1103/PhysRevA.101.053629} {\bibfield  {journal} {\bibinfo  {journal}
  {Phys. Rev. A}\ }\textbf {\bibinfo {volume} {101}},\ \bibinfo {pages}
  {053629} (\bibinfo {year} {2020})}\BibitemShut {NoStop}%
\bibitem [{\citenamefont {Mateo}\ and\ \citenamefont
  {Brand}(2015)}]{mateo2015stability}%
  \BibitemOpen
  \bibfield  {author} {\bibinfo {author} {\bibfnamefont {A.~M.}\ \bibnamefont
  {Mateo}}\ and\ \bibinfo {author} {\bibfnamefont {J.}~\bibnamefont {Brand}},\
  }\bibfield  {title} {\enquote {\bibinfo {title} {{Stability and dispersion
  relations of three-dimensional solitary waves in trapped Bose--Einstein
  condensates}},}\ }\href@noop {} {\bibfield  {journal} {\bibinfo  {journal}
  {New Journal of Physics}\ }\textbf {\bibinfo {volume} {17}},\ \bibinfo
  {pages} {125013} (\bibinfo {year} {2015})}\BibitemShut {NoStop}%
\bibitem [{\citenamefont {Anderson}\ \emph {et~al.}(2001)\citenamefont
  {Anderson}, \citenamefont {Haljan}, \citenamefont {Regal}, \citenamefont
  {Feder}, \citenamefont {Collins}, \citenamefont {Clark},\ and\ \citenamefont
  {Cornell}}]{PhysRevLett.86.2926}%
  \BibitemOpen
  \bibfield  {author} {\bibinfo {author} {\bibfnamefont {B.~P.}\ \bibnamefont
  {Anderson}}, \bibinfo {author} {\bibfnamefont {P.~C.}\ \bibnamefont
  {Haljan}}, \bibinfo {author} {\bibfnamefont {C.~A.}\ \bibnamefont {Regal}},
  \bibinfo {author} {\bibfnamefont {D.~L.}\ \bibnamefont {Feder}}, \bibinfo
  {author} {\bibfnamefont {L.~A.}\ \bibnamefont {Collins}}, \bibinfo {author}
  {\bibfnamefont {C.~W.}\ \bibnamefont {Clark}}, \ and\ \bibinfo {author}
  {\bibfnamefont {E.~A.}\ \bibnamefont {Cornell}},\ }\bibfield  {title}
  {\enquote {\bibinfo {title} {Watching dark solitons decay into vortex rings
  in a {Bose}-{Einstein} condensate},}\ }\href {\doibase
  10.1103/PhysRevLett.86.2926} {\bibfield  {journal} {\bibinfo  {journal}
  {Phys. Rev. Lett.}\ }\textbf {\bibinfo {volume} {86}},\ \bibinfo {pages}
  {2926} (\bibinfo {year} {2001})}\BibitemShut {NoStop}%
\bibitem [{\citenamefont {Shomroni}\ \emph {et~al.}(2009)\citenamefont
  {Shomroni}, \citenamefont {Lahoud}, \citenamefont {Levy},\ and\ \citenamefont
  {Steinhauer}}]{Shomroni2009}%
  \BibitemOpen
  \bibfield  {author} {\bibinfo {author} {\bibfnamefont {I.}~\bibnamefont
  {Shomroni}}, \bibinfo {author} {\bibfnamefont {E.}~\bibnamefont {Lahoud}},
  \bibinfo {author} {\bibfnamefont {S.}~\bibnamefont {Levy}}, \ and\ \bibinfo
  {author} {\bibfnamefont {J.}~\bibnamefont {Steinhauer}},\ }\bibfield  {title}
  {\enquote {\bibinfo {title} {Evidence for an oscillating soliton/vortex ring
  by density engineering of a {Bose}-{Einstein} condensate},}\ }\href {\doibase
  10.1038/nphys1177} {\bibfield  {journal} {\bibinfo  {journal} {Nature
  Physics}\ }\textbf {\bibinfo {volume} {5}},\ \bibinfo {pages} {193} (\bibinfo
  {year} {2009})}\BibitemShut {NoStop}%
\bibitem [{\citenamefont {Ku}\ \emph {et~al.}(2014)\citenamefont {Ku},
  \citenamefont {Ji}, \citenamefont {Mukherjee}, \citenamefont
  {Guardado-Sanchez}, \citenamefont {Cheuk}, \citenamefont {Yefsah},\ and\
  \citenamefont {Zwierlein}}]{PhysRevLett.113.065301}%
  \BibitemOpen
  \bibfield  {author} {\bibinfo {author} {\bibfnamefont {M.~J.~H.}\
  \bibnamefont {Ku}}, \bibinfo {author} {\bibfnamefont {W.}~\bibnamefont {Ji}},
  \bibinfo {author} {\bibfnamefont {B.}~\bibnamefont {Mukherjee}}, \bibinfo
  {author} {\bibfnamefont {E.}~\bibnamefont {Guardado-Sanchez}}, \bibinfo
  {author} {\bibfnamefont {L.~W.}\ \bibnamefont {Cheuk}}, \bibinfo {author}
  {\bibfnamefont {T.}~\bibnamefont {Yefsah}}, \ and\ \bibinfo {author}
  {\bibfnamefont {M.~W.}\ \bibnamefont {Zwierlein}},\ }\bibfield  {title}
  {\enquote {\bibinfo {title} {{Motion of a Solitonic Vortex in the BEC-BCS
  Crossover}},}\ }\href {\doibase 10.1103/PhysRevLett.113.065301} {\bibfield
  {journal} {\bibinfo  {journal} {Phys. Rev. Lett.}\ }\textbf {\bibinfo
  {volume} {113}},\ \bibinfo {pages} {065301} (\bibinfo {year}
  {2014})}\BibitemShut {NoStop}%
\bibitem [{\citenamefont {Negretti}\ and\ \citenamefont
  {Henkel}(2004)}]{Negretti_2004}%
  \BibitemOpen
  \bibfield  {author} {\bibinfo {author} {\bibfnamefont {A.}~\bibnamefont
  {Negretti}}\ and\ \bibinfo {author} {\bibfnamefont {C.}~\bibnamefont
  {Henkel}},\ }\bibfield  {title} {\enquote {\bibinfo {title} {Enhanced phase
  sensitivity and soliton formation in an integrated {BEC} interferometer},}\
  }\href {\doibase 10.1088/0953-4075/37/23/l02} {\bibfield  {journal} {\bibinfo
   {journal} {Journal of Physics B: Atomic, Molecular and Optical Physics}\
  }\textbf {\bibinfo {volume} {37}},\ \bibinfo {pages} {L385} (\bibinfo {year}
  {2004})}\BibitemShut {NoStop}%
\bibitem [{\citenamefont {Shaukat}\ \emph {et~al.}(2017)\citenamefont
  {Shaukat}, \citenamefont {Castro},\ and\ \citenamefont
  {Ter\ifmmode~\mbox{\c{c}}\else \c{c}\fi{}as}}]{PhysRevA.95.053618}%
  \BibitemOpen
  \bibfield  {author} {\bibinfo {author} {\bibfnamefont {M.~I.}\ \bibnamefont
  {Shaukat}}, \bibinfo {author} {\bibfnamefont {E.~V.}\ \bibnamefont {Castro}},
  \ and\ \bibinfo {author} {\bibfnamefont {H.}~\bibnamefont
  {Ter\ifmmode~\mbox{\c{c}}\else \c{c}\fi{}as}},\ }\bibfield  {title} {\enquote
  {\bibinfo {title} {Quantum dark solitons as qubits in {Bose}-{Einstein}
  condensates},}\ }\href {\doibase 10.1103/PhysRevA.95.053618} {\bibfield
  {journal} {\bibinfo  {journal} {Phys. Rev. A}\ }\textbf {\bibinfo {volume}
  {95}},\ \bibinfo {pages} {053618} (\bibinfo {year} {2017})}\BibitemShut
  {NoStop}%
\bibitem [{\citenamefont {Komineas}(2007)}]{komineas}%
  \BibitemOpen
  \bibfield  {author} {\bibinfo {author} {\bibfnamefont {S.}~\bibnamefont
  {Komineas}},\ }\bibfield  {title} {\enquote {\bibinfo {title} {Vortex rings
  and solitary waves in trapped {Bose}--{Einstein} condensates},}\ }\href
  {\doibase 10.1140/epjst/e2007-00206-8} {\bibfield  {journal} {\bibinfo
  {journal} {Eur. Phys. J.-Spec. Top.}\ }\textbf {\bibinfo {volume} {147}},\
  \bibinfo {pages} {133} (\bibinfo {year} {2007})}\BibitemShut {NoStop}%
\bibitem [{\citenamefont {Komineas}\ and\ \citenamefont
  {Papanicolaou}(2003)}]{PhysRevA.68.043617}%
  \BibitemOpen
  \bibfield  {author} {\bibinfo {author} {\bibfnamefont {S.}~\bibnamefont
  {Komineas}}\ and\ \bibinfo {author} {\bibfnamefont {N.}~\bibnamefont
  {Papanicolaou}},\ }\bibfield  {title} {\enquote {\bibinfo {title} {Solitons,
  solitonic vortices, and vortex rings in a confined {Bose}-{Einstein}
  condensate},}\ }\href {\doibase 10.1103/PhysRevA.68.043617} {\bibfield
  {journal} {\bibinfo  {journal} {Phys. Rev. A}\ }\textbf {\bibinfo {volume}
  {68}},\ \bibinfo {pages} {043617} (\bibinfo {year} {2003})}\BibitemShut
  {NoStop}%
\bibitem [{\citenamefont {Brand}\ and\ \citenamefont
  {Reinhardt}(2002)}]{PhysRevA.65.043612}%
  \BibitemOpen
  \bibfield  {author} {\bibinfo {author} {\bibfnamefont {J.}~\bibnamefont
  {Brand}}\ and\ \bibinfo {author} {\bibfnamefont {W.~P.}\ \bibnamefont
  {Reinhardt}},\ }\bibfield  {title} {\enquote {\bibinfo {title} {Solitonic
  vortices and the fundamental modes of the ``snake instability'': Possibility
  of observation in the gaseous {Bose}-{Einstein} condensate},}\ }\href
  {\doibase 10.1103/PhysRevA.65.043612} {\bibfield  {journal} {\bibinfo
  {journal} {Phys. Rev. A}\ }\textbf {\bibinfo {volume} {65}},\ \bibinfo
  {pages} {043612} (\bibinfo {year} {2002})}\BibitemShut {NoStop}%
\bibitem [{\citenamefont {Becker}\ \emph {et~al.}(2013)\citenamefont {Becker},
  \citenamefont {Sengstock}, \citenamefont {Schmelcher}, \citenamefont
  {Kevrekidis},\ and\ \citenamefont {Carretero-Gonz{\'{a}}lez}}]{Becker_2013}%
  \BibitemOpen
  \bibfield  {author} {\bibinfo {author} {\bibfnamefont {C.}~\bibnamefont
  {Becker}}, \bibinfo {author} {\bibfnamefont {K.}~\bibnamefont {Sengstock}},
  \bibinfo {author} {\bibfnamefont {P.}~\bibnamefont {Schmelcher}}, \bibinfo
  {author} {\bibfnamefont {P.~G.}\ \bibnamefont {Kevrekidis}}, \ and\ \bibinfo
  {author} {\bibfnamefont {R.}~\bibnamefont {Carretero-Gonz{\'{a}}lez}},\
  }\bibfield  {title} {\enquote {\bibinfo {title} {Inelastic collisions of
  solitary waves in anisotropic {Bose}-{Einstein} condensates: sling-shot
  events and expanding collision bubbles},}\ }\href {\doibase
  10.1088/1367-2630/15/11/113028} {\bibfield  {journal} {\bibinfo  {journal}
  {New Journal of Physics}\ }\textbf {\bibinfo {volume} {15}},\ \bibinfo
  {pages} {113028} (\bibinfo {year} {2013})}\BibitemShut {NoStop}%
\bibitem [{\citenamefont {Tylutki}\ \emph {et~al.}(2015)\citenamefont
  {Tylutki}, \citenamefont {Donadello}, \citenamefont {Serafini}, \citenamefont
  {Pitaevskii}, \citenamefont {Dalfovo}, \citenamefont {Lamporesi},\ and\
  \citenamefont {Ferrari}}]{tylutki2015solitonic}%
  \BibitemOpen
  \bibfield  {author} {\bibinfo {author} {\bibfnamefont {M.}~\bibnamefont
  {Tylutki}}, \bibinfo {author} {\bibfnamefont {S.}~\bibnamefont {Donadello}},
  \bibinfo {author} {\bibfnamefont {S.}~\bibnamefont {Serafini}}, \bibinfo
  {author} {\bibfnamefont {L.~P.}\ \bibnamefont {Pitaevskii}}, \bibinfo
  {author} {\bibfnamefont {F.}~\bibnamefont {Dalfovo}}, \bibinfo {author}
  {\bibfnamefont {G.}~\bibnamefont {Lamporesi}}, \ and\ \bibinfo {author}
  {\bibfnamefont {G.}~\bibnamefont {Ferrari}},\ }\bibfield  {title} {\enquote
  {\bibinfo {title} {Solitonic vortices in {Bose}-{Einstein} condensates},}\
  }\href@noop {} {\bibfield  {journal} {\bibinfo  {journal} {The European
  Physical Journal Special Topics}\ }\textbf {\bibinfo {volume} {224}},\
  \bibinfo {pages} {577} (\bibinfo {year} {2015})}\BibitemShut {NoStop}%
\bibitem [{\citenamefont {Tsatsos}\ \emph {et~al.}(2016)\citenamefont
  {Tsatsos}, \citenamefont {Edmonds},\ and\ \citenamefont
  {Parker}}]{PhysRevA.94.023627}%
  \BibitemOpen
  \bibfield  {author} {\bibinfo {author} {\bibfnamefont {M.~C.}\ \bibnamefont
  {Tsatsos}}, \bibinfo {author} {\bibfnamefont {M.~J.}\ \bibnamefont
  {Edmonds}}, \ and\ \bibinfo {author} {\bibfnamefont {N.~G.}\ \bibnamefont
  {Parker}},\ }\bibfield  {title} {\enquote {\bibinfo {title} {Transition from
  vortices to solitonic vortices in trapped atomic {Bose}-{Einstein}
  condensates},}\ }\href {\doibase 10.1103/PhysRevA.94.023627} {\bibfield
  {journal} {\bibinfo  {journal} {Phys. Rev. A}\ }\textbf {\bibinfo {volume}
  {94}},\ \bibinfo {pages} {023627} (\bibinfo {year} {2016})}\BibitemShut
  {NoStop}%
\bibitem [{\citenamefont {Donadello}\ \emph {et~al.}(2014)\citenamefont
  {Donadello}, \citenamefont {Serafini}, \citenamefont {Tylutki}, \citenamefont
  {Pitaevskii}, \citenamefont {Dalfovo}, \citenamefont {Lamporesi},\ and\
  \citenamefont {Ferrari}}]{PhysRevLett.113.065302}%
  \BibitemOpen
  \bibfield  {author} {\bibinfo {author} {\bibfnamefont {S.}~\bibnamefont
  {Donadello}}, \bibinfo {author} {\bibfnamefont {S.}~\bibnamefont {Serafini}},
  \bibinfo {author} {\bibfnamefont {M.}~\bibnamefont {Tylutki}}, \bibinfo
  {author} {\bibfnamefont {L.~P.}\ \bibnamefont {Pitaevskii}}, \bibinfo
  {author} {\bibfnamefont {F.}~\bibnamefont {Dalfovo}}, \bibinfo {author}
  {\bibfnamefont {G.}~\bibnamefont {Lamporesi}}, \ and\ \bibinfo {author}
  {\bibfnamefont {G.}~\bibnamefont {Ferrari}},\ }\bibfield  {title} {\enquote
  {\bibinfo {title} {Observation of solitonic vortices in {Bose}-{Einstein}
  condensates},}\ }\href {\doibase 10.1103/PhysRevLett.113.065302} {\bibfield
  {journal} {\bibinfo  {journal} {Phys. Rev. Lett.}\ }\textbf {\bibinfo
  {volume} {113}},\ \bibinfo {pages} {065302} (\bibinfo {year}
  {2014})}\BibitemShut {NoStop}%
\bibitem [{\citenamefont {Muryshev}\ \emph {et~al.}(2002)\citenamefont
  {Muryshev}, \citenamefont {Shlyapnikov}, \citenamefont {Ertmer},
  \citenamefont {Sengstock},\ and\ \citenamefont
  {Lewenstein}}]{PhysRevLett.89.110401}%
  \BibitemOpen
  \bibfield  {author} {\bibinfo {author} {\bibfnamefont {A.}~\bibnamefont
  {Muryshev}}, \bibinfo {author} {\bibfnamefont {G.~V.}\ \bibnamefont
  {Shlyapnikov}}, \bibinfo {author} {\bibfnamefont {W.}~\bibnamefont {Ertmer}},
  \bibinfo {author} {\bibfnamefont {K.}~\bibnamefont {Sengstock}}, \ and\
  \bibinfo {author} {\bibfnamefont {M.}~\bibnamefont {Lewenstein}},\ }\bibfield
   {title} {\enquote {\bibinfo {title} {Dynamics of dark solitons in elongated
  {Bose}-{Einstein} condensates},}\ }\href {\doibase
  10.1103/PhysRevLett.89.110401} {\bibfield  {journal} {\bibinfo  {journal}
  {Phys. Rev. Lett.}\ }\textbf {\bibinfo {volume} {89}},\ \bibinfo {pages}
  {110401} (\bibinfo {year} {2002})}\BibitemShut {NoStop}%
\bibitem [{sup()}]{supple}%
  \BibitemOpen
  \href@noop {} {}\bibinfo {note} {See Supplemental Material at
  [\url{URL_will_be_inserted_by_publisher}] for corresponding GPE
  dynamics}\BibitemShut {NoStop}%
\bibitem [{\citenamefont {Ticknor}\ \emph {et~al.}(2018)\citenamefont
  {Ticknor}, \citenamefont {Wang},\ and\ \citenamefont
  {Kevrekidis}}]{PhysRevA.98.033609}%
  \BibitemOpen
  \bibfield  {author} {\bibinfo {author} {\bibfnamefont {C.}~\bibnamefont
  {Ticknor}}, \bibinfo {author} {\bibfnamefont {W.}~\bibnamefont {Wang}}, \
  and\ \bibinfo {author} {\bibfnamefont {P.~G.}\ \bibnamefont {Kevrekidis}},\
  }\bibfield  {title} {\enquote {\bibinfo {title} {Spectral and dynamical
  analysis of a single vortex ring in anisotropic harmonically trapped
  three-dimensional {Bose}-{Einstein} condensates},}\ }\href {\doibase
  10.1103/PhysRevA.98.033609} {\bibfield  {journal} {\bibinfo  {journal} {Phys.
  Rev. A}\ }\textbf {\bibinfo {volume} {98}},\ \bibinfo {pages} {033609}
  (\bibinfo {year} {2018})}\BibitemShut {NoStop}%
\bibitem [{\citenamefont {Cockburn}\ \emph {et~al.}(2010)\citenamefont
  {Cockburn}, \citenamefont {Nistazakis}, \citenamefont {Horikis},
  \citenamefont {Kevrekidis}, \citenamefont {Proukakis},\ and\ \citenamefont
  {Frantzeskakis}}]{prouka}%
  \BibitemOpen
  \bibfield  {author} {\bibinfo {author} {\bibfnamefont {S.~P.}\ \bibnamefont
  {Cockburn}}, \bibinfo {author} {\bibfnamefont {H.~E.}\ \bibnamefont
  {Nistazakis}}, \bibinfo {author} {\bibfnamefont {T.~P.}\ \bibnamefont
  {Horikis}}, \bibinfo {author} {\bibfnamefont {P.~G.}\ \bibnamefont
  {Kevrekidis}}, \bibinfo {author} {\bibfnamefont {N.~P.}\ \bibnamefont
  {Proukakis}}, \ and\ \bibinfo {author} {\bibfnamefont {D.~J.}\ \bibnamefont
  {Frantzeskakis}},\ }\bibfield  {title} {\enquote {\bibinfo {title}
  {Matter-wave dark solitons: Stochastic versus analytical results},}\ }\href
  {\doibase 10.1103/PhysRevLett.104.174101} {\bibfield  {journal} {\bibinfo
  {journal} {Phys. Rev. Lett.}\ }\textbf {\bibinfo {volume} {104}},\ \bibinfo
  {pages} {174101} (\bibinfo {year} {2010})}\BibitemShut {NoStop}%
\bibitem [{\citenamefont {Busch}\ and\ \citenamefont
  {Anglin}(2000)}]{PhysRevLett.84.2298}%
  \BibitemOpen
  \bibfield  {author} {\bibinfo {author} {\bibfnamefont {T.}~\bibnamefont
  {Busch}}\ and\ \bibinfo {author} {\bibfnamefont {J.~R.}\ \bibnamefont
  {Anglin}},\ }\bibfield  {title} {\enquote {\bibinfo {title} {Motion of dark
  solitons in trapped {Bose}-{Einstein} condensates},}\ }\href {\doibase
  10.1103/PhysRevLett.84.2298} {\bibfield  {journal} {\bibinfo  {journal}
  {Phys. Rev. Lett.}\ }\textbf {\bibinfo {volume} {84}},\ \bibinfo {pages}
  {2298} (\bibinfo {year} {2000})}\BibitemShut {NoStop}%
\bibitem [{\citenamefont {Li}\ \emph {et~al.}(2008)\citenamefont {Li},
  \citenamefont {Haque},\ and\ \citenamefont {Komineas}}]{PhysRevA.77.053610}%
  \BibitemOpen
  \bibfield  {author} {\bibinfo {author} {\bibfnamefont {W.}~\bibnamefont
  {Li}}, \bibinfo {author} {\bibfnamefont {M.}~\bibnamefont {Haque}}, \ and\
  \bibinfo {author} {\bibfnamefont {S.}~\bibnamefont {Komineas}},\ }\bibfield
  {title} {\enquote {\bibinfo {title} {Vortex dipole in a trapped
  two-dimensional {Bose}-{Einstein} condensate},}\ }\href {\doibase
  10.1103/PhysRevA.77.053610} {\bibfield  {journal} {\bibinfo  {journal} {Phys.
  Rev. A}\ }\textbf {\bibinfo {volume} {77}},\ \bibinfo {pages} {053610}
  (\bibinfo {year} {2008})}\BibitemShut {NoStop}%
\bibitem [{\citenamefont {Middelkamp}\ \emph {et~al.}(2010)\citenamefont
  {Middelkamp}, \citenamefont {Kevrekidis}, \citenamefont {Frantzeskakis},
  \citenamefont {Carretero-Gonz\'alez},\ and\ \citenamefont
  {Schmelcher}}]{PhysRevA.82.013646}%
  \BibitemOpen
  \bibfield  {author} {\bibinfo {author} {\bibfnamefont {S.}~\bibnamefont
  {Middelkamp}}, \bibinfo {author} {\bibfnamefont {P.~G.}\ \bibnamefont
  {Kevrekidis}}, \bibinfo {author} {\bibfnamefont {D.~J.}\ \bibnamefont
  {Frantzeskakis}}, \bibinfo {author} {\bibfnamefont {R.}~\bibnamefont
  {Carretero-Gonz\'alez}}, \ and\ \bibinfo {author} {\bibfnamefont
  {P.}~\bibnamefont {Schmelcher}},\ }\bibfield  {title} {\enquote {\bibinfo
  {title} {Bifurcations, stability, and dynamics of multiple matter-wave vortex
  states},}\ }\href {\doibase 10.1103/PhysRevA.82.013646} {\bibfield  {journal}
  {\bibinfo  {journal} {Phys. Rev. A}\ }\textbf {\bibinfo {volume} {82}},\
  \bibinfo {pages} {013646} (\bibinfo {year} {2010})}\BibitemShut {NoStop}%
\bibitem [{\citenamefont {Mithun}\ \emph {et~al.}(2022)\citenamefont {Mithun},
  \citenamefont {Carretero-Gonz\'alez}, \citenamefont {Charalampidis},
  \citenamefont {Hall},\ and\ \citenamefont
  {Kevrekidis}}]{PhysRevA.105.053303}%
  \BibitemOpen
  \bibfield  {author} {\bibinfo {author} {\bibfnamefont {T.}~\bibnamefont
  {Mithun}}, \bibinfo {author} {\bibfnamefont {R.}~\bibnamefont
  {Carretero-Gonz\'alez}}, \bibinfo {author} {\bibfnamefont {E.~G.}\
  \bibnamefont {Charalampidis}}, \bibinfo {author} {\bibfnamefont {D.~S.}\
  \bibnamefont {Hall}}, \ and\ \bibinfo {author} {\bibfnamefont {P.~G.}\
  \bibnamefont {Kevrekidis}},\ }\bibfield  {title} {\enquote {\bibinfo {title}
  {Existence, stability, and dynamics of monopole and alice ring solutions in
  antiferromagnetic spinor condensates},}\ }\href {\doibase
  10.1103/PhysRevA.105.053303} {\bibfield  {journal} {\bibinfo  {journal}
  {Phys. Rev. A}\ }\textbf {\bibinfo {volume} {105}},\ \bibinfo {pages}
  {053303} (\bibinfo {year} {2022})}\BibitemShut {NoStop}%
\bibitem [{\citenamefont {Gaunt}\ \emph {et~al.}(2013)\citenamefont {Gaunt},
  \citenamefont {Schmidutz}, \citenamefont {Gotlibovych}, \citenamefont
  {Smith},\ and\ \citenamefont {Hadzibabic}}]{PhysRevLett.110.200406}%
  \BibitemOpen
  \bibfield  {author} {\bibinfo {author} {\bibfnamefont {A.~L.}\ \bibnamefont
  {Gaunt}}, \bibinfo {author} {\bibfnamefont {T.~F.}\ \bibnamefont
  {Schmidutz}}, \bibinfo {author} {\bibfnamefont {I.}~\bibnamefont
  {Gotlibovych}}, \bibinfo {author} {\bibfnamefont {R.~P.}\ \bibnamefont
  {Smith}}, \ and\ \bibinfo {author} {\bibfnamefont {Z.}~\bibnamefont
  {Hadzibabic}},\ }\bibfield  {title} {\enquote {\bibinfo {title}
  {{Bose}-{Einstein} condensation of atoms in a uniform potential},}\ }\href
  {\doibase 10.1103/PhysRevLett.110.200406} {\bibfield  {journal} {\bibinfo
  {journal} {Phys. Rev. Lett.}\ }\textbf {\bibinfo {volume} {110}},\ \bibinfo
  {pages} {200406} (\bibinfo {year} {2013})}\BibitemShut {NoStop}%
\bibitem [{\citenamefont {Tsitoura}\ \emph {et~al.}(2016)\citenamefont
  {Tsitoura}, \citenamefont {Anastassi}, \citenamefont {Marzuola},
  \citenamefont {Kevrekidis},\ and\ \citenamefont
  {Frantzeskakis}}]{PhysRevA.94.063612}%
  \BibitemOpen
  \bibfield  {author} {\bibinfo {author} {\bibfnamefont {F.}~\bibnamefont
  {Tsitoura}}, \bibinfo {author} {\bibfnamefont {Z.~A.}\ \bibnamefont
  {Anastassi}}, \bibinfo {author} {\bibfnamefont {J.~L.}\ \bibnamefont
  {Marzuola}}, \bibinfo {author} {\bibfnamefont {P.~G.}\ \bibnamefont
  {Kevrekidis}}, \ and\ \bibinfo {author} {\bibfnamefont {D.~J.}\ \bibnamefont
  {Frantzeskakis}},\ }\bibfield  {title} {\enquote {\bibinfo {title} {Dark
  solitons near potential and nonlinearity steps},}\ }\href {\doibase
  10.1103/PhysRevA.94.063612} {\bibfield  {journal} {\bibinfo  {journal} {Phys.
  Rev. A}\ }\textbf {\bibinfo {volume} {94}},\ \bibinfo {pages} {063612}
  (\bibinfo {year} {2016})}\BibitemShut {NoStop}%
\bibitem [{\citenamefont {Aycock}\ \emph {et~al.}(2017)\citenamefont {Aycock},
  \citenamefont {Hurst}, \citenamefont {Efimkin}, \citenamefont {Genkina},
  \citenamefont {Lu}, \citenamefont {Galitski},\ and\ \citenamefont
  {Spielman}}]{doi:10.1073/pnas.1615004114}%
  \BibitemOpen
  \bibfield  {author} {\bibinfo {author} {\bibfnamefont {L.~M.}\ \bibnamefont
  {Aycock}}, \bibinfo {author} {\bibfnamefont {H.~M.}\ \bibnamefont {Hurst}},
  \bibinfo {author} {\bibfnamefont {D.~K.}\ \bibnamefont {Efimkin}}, \bibinfo
  {author} {\bibfnamefont {D.}~\bibnamefont {Genkina}}, \bibinfo {author}
  {\bibfnamefont {H.-I.}\ \bibnamefont {Lu}}, \bibinfo {author} {\bibfnamefont
  {V.~M.}\ \bibnamefont {Galitski}}, \ and\ \bibinfo {author} {\bibfnamefont
  {I.~B.}\ \bibnamefont {Spielman}},\ }\bibfield  {title} {\enquote {\bibinfo
  {title} {Brownian motion of solitons in a {Bose}-{Einstein} condensate},}\
  }\href {\doibase 10.1073/pnas.1615004114} {\bibfield  {journal} {\bibinfo
  {journal} {Proceedings of the National Academy of Sciences}\ }\textbf
  {\bibinfo {volume} {114}},\ \bibinfo {pages} {2503} (\bibinfo {year}
  {2017})}\BibitemShut {NoStop}%
\end{thebibliography}%


\providecommand{\noopsort}[1]{}\providecommand{\singleletter}[1]{#1}%
%
% \clearpage
% \newpage

%%%%%%%%%%%%%%
%######################################
\end{document}